\documentclass[12pt]{article}

\RequirePackage{amsmath, amsthm, amssymb, amsfonts, booktabs,
  graphicx, natbib, setspace}
\usepackage{xcolor}

\usepackage{mathtools}

\allowdisplaybreaks

\graphicspath{{./figs/}}

\DeclareMathOperator{\Var}{\mathbb{V}}
\DeclareMathOperator{\Tr}{Tr}
\DeclareMathOperator{\diag}{diag}
\DeclareMathOperator{\sgn}{sgn}

\newtheorem{theorem}{Theorem}
\newtheorem{lemma}{Lemma}

\newtheorem{assumption}{Assumption}
\newtheorem{remark}{Remark}

\usepackage[left=1in,top=1in,right=1in,bottom=1in]{geometry}

\usepackage[colorlinks,linkcolor=blue,citecolor=blue,urlcolor=blue,filecolor=blue]{hyperref}

\allowdisplaybreaks

\newcommand*{\email}[1]{%
    \href{mailto:#1}{\color{black}{#1}}\par
    }

\begin{document}

\date{\vspace{-5ex}}

\title{Regularized Fingerprinting in Detection and Attribution of
  Climate Change with Weight Matrix Optimizing the Efficiency in
  Scaling Factor Estimation}

\author{{Yan Li$^{1}$}\thanks{Corresponding author. Email: \email{yan.4.li@uconn.edu}}, Kun Chen$^{1}$, Jun Yan$^{1}$, and Xuebin Zhang$^{2}$ \\[1.5ex]
  $^{1}$Department of Statistics, University of Connecticut, CT, U.S.A. \\
  $^{2}$Environment and Climate Change Canada, QC, CA}

\maketitle





\begin{abstract}
  The optimal fingerprinting method for detection and attribution
  of climate change is based on a multiple regression where each
  covariate has measurement error whose covariance matrix is the same
  as that of the regression error up to a known scale. Inferences
  about the regression coefficients are critical not only for making
  statements about detection and attribution but also for quantifying
  the uncertainty in important outcomes derived from detection and
  attribution analyses. When there is no errors-in-variables (EIV),
  the optimal weight matrix in estimating
  the regression coefficients is the precision matrix of the
  regression error which, in practice, is never known and has to be
  estimated from climate model simulations. We construct a weight
  matrix by inverting a nonlinear shrinkage estimate of the error
  covariance matrix that minimizes loss functions directly
  targeting the uncertainty of the resulting regression coefficient
  estimator. The resulting estimator of the regression coefficients is
  asymptotically optimal as the sample size of the climate
  model simulations and the matrix dimension go to infinity together
  with a limiting ratio. When EIVs are present, the estimator of the
  regression coefficients based on the proposed weight matrix
  is asymptotically more efficient than that based on the inverse of
  the existing linear shrinkage estimator of the error covariance
  matrix. The performance of the method is confirmed in finite sample
  simulation studies mimicking realistic situations in terms of the
  length of the confidence intervals and empirical coverage rates for
  the regression coefficients. An application to detection and
  attribution analyses of the mean temperature at different spatial
  scales illustrates the utility of the method.
\end{abstract}
\vfill

\newpage
\doublespacing

\section{Introduction}
\label{sec:intro}

Detection and attribution analyses of climate change are critical
components in establishing a causal relationship from the human
emission of greenhouse gases to the warming of planet Earth
\citep[e.g.,][]{Bind:etal:dete:2013}. In climate science, detection is the
process of demonstrating that a climate variable has changed in some
defined statistical sense without providing a reason for that change;
attribution is the process of evaluating the relative contributions of
multiple causal factors to a change or event with an assignment of
statistical confidence \citep[e.g.,][]{Hege:Zwie:use:2011}. Casual
factors usually refer to external forcings, which may be anthropogenic
(e.g., greenhouse gases, aerosols, ozone precursors, land use) and/or
natural (e.g., volcanic eruptions, solar cycle modulations). By comparing
simulated results of climate models with observed climate variables, a
detection and attribution analysis evaluates the consistency of
observed changes with the expected response, also known as
fingerprint, of the climate system under each external forcing.

Optimal fingerprinting is the most widely used method for detection
and attribution analyses \citep[e.g.,][]{Hass:mult:1997,
  Alle:Stot:esti:2003, Hege:etal:good:2010}.
Fingerprinting is a procedure that regresses
the observed climate variable of interest on the fingerprints of
external forcings, and checks whether the fingerprints are found in
and consistent with the observed data. The
central target of statistical inferences here is the regression
coefficients, also known as scaling factors, which scale the
fingerprints of external forcings to best match the observed climate
change. Historically, the method was ``optimal'' in the context
of generalized least squares (GLS) when the precision
matrix of the regression error is used as weight, such that the
resulting estimator of the scaling factors have the smallest
variances. It was later recognized that the fingerprint covariates are
not observed but estimated from climate model simulations. This leads to
an errors-in-variables (EIV) issue, which has been approached by the total least
squares (TLS) \citep{Alle:Stot:esti:2003} with both the response and
the covariates ``prewhittened'' by the covariance matrix of the error.
The covariance matrix represents the internal climate variation. In
practice,  it is unknown and has to be estimated
from climate model simulations \citep{Alle:Tett:chec:1999,
  Ribe:Plan:Terr:appl:2013}, which is generally handled preliminarily
and independently from the regression inference
\citep[e.g.,][]{hannart2014optimal}.

Estimating the error covariance matrix in fingerprinting
is challenging because the available runs from climate model
simulations are small relative to the dimension of the covariance
matrix. The dimension of the covariate matrix is often reduced
by considering averages over 5-year or 10-year blocks in time and/or
over aggregated grid-boxes in space.
For example, decadal average over a 110-year
period in 25 grid-boxes would lead to a dimension of $275\times 275$.
The sample size from climate model simulations that can be used is,
however, at most a few hundreds. The sample covariance matrix is not
invertible when the sample size of climate model simulations is
less than its dimension. Earlier methods project data onto the leading
empirical orthogonal functions (EOF) of the internal climate variability
\citep{hegerl1996detecting, Alle:Tett:chec:1999}. More recently, the
regularized optimal fingerprinting (ROF) method avoids the projection
step by a regularized covariance matrix estimation
\citep{ribes2009adaptation}, which is based on the linear shrinkage
covariance matrix estimator of \citet{ledoit2004well}. The ROF method
has been shown to provide a more accurate implementation of optimal
fingerprinting than the EOF projection method
\citep{Ribe:Plan:Terr:appl:2013}.

The uncertainty in estimating the error covariance matrix has
important implications in optimal fingerprinting. The optimality in
inferences about the scaling factor in optimal fingerprinting was
historically based on the assumption that the error covariance matrix
is known. The properties of the scaling factor estimator obtained by
substituting the error covariance matrix with an estimate have not
been thoroughly investigated in the literature. For example,
it is not until recently that the confidence intervals for
the scaling factors constructed from asymptotic normal approximation
\citep{fuller1980properties, gleser1981estimation} or
bootstrap \citep{pesta2012total} were reported to be overly narrow when
the matrix is only known up to a scale \citep{delsole2019} or
completely unknown \citep{li2019confidence}. A natural, fundamental
question is: when the error covariance matrix is estimated, are the
confidence intervals constructed using the optimal approach under the
assumption of known error covariance matrix still optimal? Since
optimality under unknown error covariance matrix is practically
infeasible, to be precise, we term fingerprinting for optimal
fingerprinting and regularized fingerprinting (RF) for ROF in the
sequel.

The contributions of this paper are two-fold. First, we develop a new
method to construct the weight matrix in RF by minimizing directly
the uncertainty of the resulting scaling factor estimators. The weight
matrix is the inverse of a nonlinear shrinkage estimator of the error
covariance matrix inspired by \citet{ledoit2017direct}. We first
extend the validity of their nonlinear shrinkage estimator
to the context of RF via GLS regression with no EIV.
We show that the proposed method is asymptotically optimal as
the sample size of climate model simulations
and the matrix dimension go to infinity together
with a fixed ratio. When there is EIV, as is the case in practice,
we show that the proposed weight is more efficient than the existing
weight in RF \citep{Ribe:Plan:Terr:appl:2013} in terms of the
asymptotic variance of the scaling factor estimator when RF is
conducted with generalized TLS (GTLS). This is why we refer to
the current practice by RF instead of ROF.
The second contribution is practical recommendations for assumptions
about the structure of the error covariance matrix under which the
sample covariance is estimated before any shrinkage in RF based on
findings of a comparison study under various realistic settings.
An implementation of both linear shrinkage and nonlinear shrinkage
estimators is publicly available in an R package \texttt{dacc}
\citep{Rpkg:dacc} for detection and attribution of climate change.

The rest of this article is organized as follows. After a review of
RF in Section~\ref{sec:rof}, we develop the proposed weight matrix
and the theoretical results to support the asymptotic performance of
proposed method in Section~\ref{sec:weight}. A large scale numerical
study assessing the performance of the proposed method is reported in
Section~\ref{sec:numeric}. In Section~\ref{sec:real}, we demonstrate
the proposed method with detection and attribution analysis on global
and regional scales. A discussion concludes in Section~\ref{sec:disc}.
The technical proofs of the theoretical results are relegated in the
Supplementary Materials.

\section{Regularized Fingerprinting}
\label{sec:rof}

Fingerprinting takes the form of a linear regression with EIV
\begin{align}
\label{eq:fp}
Y &= \sum_{i=1}^p X_i \beta_i + \epsilon,\\
\label{eq:eiv}
\tilde X_i &= X_i + \nu_i, \qquad i = 1, \ldots, p,
\end{align}
where $Y$ is a $N \times 1$ vector of the observed climate variable of
interest, $X_i$ is the true but unobserved $N \times 1$ fingerprint
vector the $i$th external forcing with scaling factor $\beta_i$,
$\epsilon$ is a $N\times 1$ vector of normally distributed regression
error with mean zero and covariance matrix $\Sigma$,
$\tilde X_i$ is an estimate of $X_i$ based on $n_i$ climate model
simulations under the $i$th external forcing, and $\nu_i$ is a
normally distributed measurement error vector with mean zero and
covariance matrix $\Sigma / n_i$, and
$\nu_i$'s are mutually independent and independent of $\epsilon$,
$i = 1, \ldots, p$. The covariance matrices of $\nu_i$'s and
$\epsilon$ only differ in scales under the assumption that the climate
models reflect the real climate variation. No intercept is present in the
regression because the response and the covariates are centered by the
same reference level. The primary target of inference
is the scaling factors
$\beta = (\beta_1, \ldots, \beta_p)^{\top}$.

The ``optimal'' in optimal fingerprinting originated from earlier
practices under two assumptions:
1) the error covariance matrix $\Sigma$ is known; and
2) $X_i$'s are observed. The GLS
estimator of $\beta$ with weight matrix $W$ is
\[
\hat\beta(W) = (X^{\top} W X)^{-1} X^{\top} W Y,
\]
where $X = (X_1, \ldots, X_p)$.  The covariance matrix of the
estimator $\hat\beta (W)$ is
\begin{align}
  \label{eq:vbeta}
\Var(\hat\beta (W) ) &=
(X^{\top} W X)^{-1} X^{\top} W \Sigma W X(X^{\top} W X)^{-1}
\end{align}
The optimal weight matrix is $W = \Sigma^{-1}$, in which case,
$\hat \beta(\Sigma^{-1})$ is the best linear unbiased estimator of
$\beta$ with covariance matrix $X^{\top} \Sigma^{-1} X$.  Since
$\Sigma$ is unknown, a feasible version of GLS uses
$W = \hat\Sigma^{-1}$, where $\hat\Sigma$ is an estimator of
$\Sigma$ obtained separately from controlled runs of climate model
simulations.

Later on it was recognized that, instead of $X_i$'s, only their
estimates $\tilde X_i$'s are observed and that using $\tilde X_i$'s in
place of $X_i$'s leads to bias in estimating $\beta$ \citep{Alle:Stot:esti:2003}.
If $\Sigma$ is given, the same structure (up to a scale $1/n_i$) of
the covariance matrices of $\nu_i$'s and $\epsilon$ allows precise
pre-whitening of both $Y$ and $\tilde X_i$'s . Then the TLS can be
applied to the pre-whitened variables. Inferences about $\beta$ can be
based on the asymptotic normal distribution of the TLS estimator of
$\beta$ \citep{gleser1981estimation} or nonparametric bootstrap
\citep{pesta2012total}, as recently studied by \citet{delsole2019}.
Similar to the GLS setting, a feasible version of the GTLS
procedure relies on an estimator of $\Sigma$.

The current practice of fingerprinting
consists of two separate steps. First, estimate $\Sigma$ from
controlled runs of climate model simulations under the assumption that
the climate models capture the internal variability of the real
climate system. Second, use this estimated matrix to
pre-whiten both the outcome and covariates in the regression
model~\eqref{eq:fp}--\eqref{eq:eiv}, and obtain the GTLS estimator of
$\beta$ on the pre-whitened data. Nonetheless, estimation of $\Sigma$
in the first step is not an easy task. The dimension of
$\Sigma$ is $N\times N$, with $N(N + 1)/2$ parameters if no structure
is imposed, which is too large for the sample size $n$ of available
climate model simulations (usually in a few hundreds at most).
The sample covariance matrix based on the $n$~runs is a start, but it
is of too much variation; when $N > n$ it is not even invertible. The
linear shrinkage method of \citet{ledoit2004well} regularizes the
sample covariance matrix $\hat\Sigma_n$ to in the form of
$\lambda \hat\Sigma_n + \rho I$, where
$\lambda$ and $\rho$ are scalar tuning parameters and $I$ is the
identity matrix.  This class of shrinkage estimators has the effect of
shrinking the set of sample eigenvalues by reducing its dispersion
around the mean, pushing up the smaller ones and pulling down the larger
ones. This estimator has been used in the current RF practice
\citep{ribes2009adaptation, Ribe:Plan:Terr:appl:2013}.

Substituting $\Sigma$ with an estimator introduces an additional
uncertainty. The impact of this uncertainty on the properties of
resulting ROF estimator has not been investigated when the whole
structure of $\Sigma$ is unknown \citep{li2019confidence}.
The optimality of the optimal fingerprinting in its
original sense is unlikely to still hold. Now that the properties of
the resulting estimator of $\beta$ depends on an estimated weight
matrix, can we choose this weight matrix estimator to minimize the
variance of the estimator of $\beta$? The recently proposed nonlinear
shrinkage estimator \citep{ledoit2017direct, ledoit2017optimal} has
high potential to outperform the linear shrinkage estimators.


\section{Weight Matrix Construction}
\label{sec:weight}

We consider constructing the weight matrix by inverting a
nonlinear shrinkage estimator of $\Sigma$ \citep{ledoit2017nonlinear}
in the fingerprinting context. New theoretical results are developed
to justify the adaptation of this nonlinear shrinkage
estimator of $\Sigma$ to minimize the uncertainty of the resulting
estimator $\hat\beta$ of $\beta$.  Assume that there are
$n$~replicates from climate model simulations (usually pre-industrial
control runs) that are independent of $Y$ and $\tilde X_i$'s.
Let $Z_{i}, Z_{2}, \ldots, Z_{n} \in \mathbb{R}^{N}$ be the centered
replicates so that the sample covariance matrix is computed as
$\hat{\Sigma}_{n} = n^{-1}\sum_{i = 1}^{n} Z_{i} Z^{\top}_{i}$.
Our strategy is to revisit the GLS setting with no EIV first and then
apply the result of the GTLS setting to the case under EIV, the same
order as the historical development.

\subsection{GLS}
Since the target of inference is $\beta$, we propose to minimize the
``total variation'' of the covariance matrix
$\Var(\hat\beta)$ of the estimated
scale factors $\hat\beta(W)$ in~\eqref{eq:vbeta} with respect to
$W = \hat\Sigma^{-1}$.  Two loss functions are considered that
measure the variation of $\hat\beta$, namely, the summation of the
variances of $\hat\beta$ (trace of $\Var(\hat\beta)$) and the general
variance of $\hat\beta$ (determinant of $\Var(\hat\beta)$), denoted
respectively as $L_1(\hat{\Sigma}, \Sigma, X)$ and
$L_2(\hat{\Sigma}, \Sigma, X)$. In particular, we have
\begin{align*}
  L_1(\hat{\Sigma}, \Sigma, X)
  = &\Tr\big((X^{\top}\hat{\Sigma}^{-1}X)^{-1}
    X^{\top}\hat{\Sigma}^{-1}\Sigma \hat{\Sigma}^{-1}X
    (X^{\top}\hat{\Sigma}^{-1}X)^{-1}\big), \\
  L_2(\hat{\Sigma}, \Sigma, X)
  = &\left(\frac{\Tr(X^{\top}X)}{pN} \right)^p
    \det\left(\frac{X^{\top}\hat{\Sigma}^{-1}\Sigma
    \hat{\Sigma}^{-1}X}{N}\right)
    \det{}^{-2}\left(\frac{X^{\top}\hat{\Sigma}^{-1}X}{N}\right),
\end{align*}
where the first loss function directly targets on the trace of
$\Var(\hat \beta(W))$ and the second loss
function is proportional to the determinant of
$\Var(\hat\beta(W))$
(up to a constant scale $\{\Tr(X^{\top}X)/p\}^p$).

The theoretical development is built on minimizing
the limiting forms of the loss functions as $n \to \infty$
and $N\to\infty$. The special case of $p = 1$ has been approached by
\citet{ledoit2017nonlinear}. We extend their result to multiple
linear regressions with $p > 1$.

\begin{lemma} \label{Lem:orth}
  The loss functions $L_1(\hat{\Sigma}, \Sigma, X)$ and
  $L_2(\hat{\Sigma}, \Sigma, X)$
  remain unchanged after orthogonalization
  of design matrix $X$ via the singular value decomposition.
\end{lemma}

The proof of Lemma~\ref{Lem:orth} is in Appendix~\ref{Sec:orthlemma}.

Lemma~\ref{Lem:orth} implies that,
without loss of generality, we only need to consider
orthogonal designs in the regression model~\eqref{eq:fp}.  In other
words, we may assume that the columns of
the design matrix $X$ are such that $X_i^{\top}X_j = 0$ for any
$i \ne j$.

Consider the minimum variance loss function
\begin{align}
  L_{\mathrm{mv}}(\hat{\Sigma}, \Sigma)
  = \frac{\Tr(\hat{\Sigma}^{-1}\Sigma
  \hat{\Sigma}^{-1})/N}{(\Tr(\hat \Sigma^{-1}) / N)^2}
  \label{eq:mvloss}
\end{align}
derived in \citet{engle2019large}. We have the following result.

\begin{theorem}
  \label{thm:equal}
  As dimension $N \to \infty$ and sample size $n \to \infty$ with
  $N / n \to c$ for a constant $c$, minimizing $\lim_{n, N \to \infty}
  L_1(\hat{\Sigma}, \Sigma, X)$ or $\lim_{n, N \to \infty}
  L_2(\hat{\Sigma}, \Sigma, X)$ is equivalent to minimizing $\lim_{n, N
    \to \infty} L_{\mathrm{mv}}(\hat{\Sigma}, \Sigma)$.
\end{theorem}

The proof for Theorem~\ref{thm:equal} is presented in Appendix~\ref{Sec:loss}.

Let $\hat \Sigma_n = \Gamma_nD_n\Gamma_n^{\top}$ be the spectral decomposition
of the sample covariance matrix $\hat \Sigma_n$, where
$D_n = \diag(\lambda_1, \ldots, \lambda_N)$ is the diagonal matrix of
the eigenvalues and $\Gamma_n$ contains the corresponding
eigenvectors. Consider the rotation invariant class of the estimators
$\hat{\Sigma} = \Gamma_n \tilde D_n \Gamma_n^{\top}$, where
$\tilde D_n = \diag\big(\delta(\lambda_1), \ldots, \delta(\lambda_N)\big)$ for a
smooth function $\delta(\cdot)$. Then, under some regularity
assumptions on the data genration mechanism
\citep[Assumptions~1-4]{ledoit2017direct},
we can get the asymptotically optimal
estimator $\hat\Sigma$ which minimizes the
limiting form of proposed two loss functions as $n \to \infty$ and
$N \to \infty$.

Let $F_N$ be the empirical cumulative distribution function of sample
eigenvalues. \citet{silverstein1995strong} showed that the limiting
form $F = \lim_{N, n \to \infty} F_N$ exists under the same
assumptions. The oracle optimal nonlinear shrinkage estimator
minimizing the limiting form of proposed loss function under general
asymptotics depends only on the derivative $f = F'$ of
$F$ and its Hilbert transform $\mathcal{H}_f$, and the
limiting ratio $c$ of $N/n$ \citep{ledoit2017direct}, with the
shrinkage form of the eigenvalues given by
\begin{equation}\label{eq:oracle}
    \delta_{\mathrm{oracle}}(\lambda_{i}) = \frac{\lambda_{i}}{[\pi
    c\lambda_{i} f (\lambda_{i})]^{2}
    + [1 - c -  \pi c \lambda_{i}
    \mathcal{H}_{f}(\lambda_{i})]^{2}}.
\end{equation}

A feasible nonlinear shrinkage estimator ({\it bona fide} counterpart
of the oracle estimator) can be based on a kernel estimator of $f$,
which is proposed and shown by \citet{ledoit2017direct} to perform as well as the
oracle estimator asymptotically.  Let $c_n=N/n$, which is a estimator
for the limiting concentration ratio $c$. The feasible nonlinear
shrinkage $\delta(\lambda_i)$, $i = 1,\ldots,N$, of the sample
eigenvalues is defined as following results for both cases of
$c_n \le 1$ and $c_n > 1$.

\paragraph{Case 1} If $c_n \le 1$, that is, the sample covariance matrix is
nonsingular, then
\begin{align*}
    \delta(\lambda_{i}) = \frac{\lambda_{i}}{[\pi
    \frac{N}{n}\lambda_{i}\tilde{f}(\lambda_{i})]^{2}
    + [1 - \frac{N}{n} -  \pi \frac{N}{n} \lambda_{i}
    \mathcal{H}_{\tilde{f}}(\lambda_{i})]^{2}},
\end{align*}
where $\tilde{f}(\cdot)$ is a kernel estimator of the limiting
sample spectral density $f$, and
$\mathcal{H}_{\tilde{f}}$ is the Hilbert transform of $\tilde f$.
Various authors adopt different
conventions to define the Hilbert transform. We
follow \citet{ledoit2017direct} and apply the same
semicircle kernel function and Hilbert transform because of the
consistency of the resulting feasible estimator. 
Specifically, we have
\begin{align*}
  \tilde{f}(\lambda_{i}) &= \frac{1}{N}\sum_{j=1}^{N}
    \frac{\sqrt{[4\lambda^{2}_{j}h_{n}^{2} -
    (\lambda_{i}-\lambda_{j})^{2}]^{+}}}
    {2\pi \lambda^{2}_{j} h^{2}_{n}},\\
  \mathcal{H}_{\tilde{f}}(\lambda_{i}) &= \frac{1}{N}
    \sum_{j=1}^{N}\frac{\sgn(\lambda_{i} -
    \lambda_{j})\sqrt{[(\lambda_{i} - \lambda _{j})^{2} -
    4\lambda^{2}_{j}h_{n}^{2}]^{+}}  - \lambda_{i} + \lambda_{j}}{2\pi
    \lambda^{2}_{j} h^{2}_{n}},
\end{align*}
where $h_n = n^{-\gamma}$ is the bandwidth of the semicircle kernel
with tuning parameter $\gamma$,
and $a^+ = \max(0, a)$. For details on the
Hilbert transform and the mathematical formulation of Hilbert
transform for commonly used kernel functions, see
\citet{bateman1954tables}.

\paragraph{Case 2} In optimal fingerprinting applications, the case of
$c_n > 1$ is more relevant because the number $n$ of controlled runs that
can be used to estimate the internal climate variation is often
limited, much less than the dimension $N$ of the problem.
If $c_n > 1$, we have $N - n$ null eigenvalues. Assume that
$(\lambda_{1}, \ldots, \lambda_{N-n}) = 0$. In this case, we only
consider the empirical cumulative distribution function
$\underline F_N$ of the nonzero $n$ eigenvalues. From
\citet{silverstein1995strong}, there existing a limiting function
$\underline F$ such that
$\lim_{N, n \to \infty}\underline F_N = \underline F$, and it admits a
continuous derivative $\underline f$. The oracle estimator in
Equation~\eqref{eq:oracle} can be written as
\[
  \delta_{\mathrm{oracle}}(\lambda_{i}) = \frac{\lambda_{i}}{\pi^2
    \lambda_{i}^2 [\underline f (\lambda_{i})^{2} +
    \mathcal{H}_{\underline f}(\lambda_{i})^{2}]}.
\]
Then the kernel approach can be adapted in this case. Let
$\tilde{\underline{f}}$ and $H_{\tilde{\underline{f}}}$ be,
respectively, the kernel estimator for $\underline f$ and its Hilbert
transform $H_{\underline f}$. The our feasible shrinkage estimator is
\begin{align*}
  \delta(0)
  &= \frac{1}{\pi \frac{N -
    n}{n}\mathcal{H}_{\tilde{\underline{f}}}(0)},
    \quad i = 1, \ldots, N-n,\\
  \delta(\lambda_{i})
  &=
    \frac{\lambda_{i}}{\pi^{2}\lambda^{2}_{i}[\tilde{\underline{f}}(\lambda_{i})^{2} +
    \mathcal{H}_{\tilde{\underline{f}}}(\lambda_{i})^{2}]},
    \quad i = N-n+1, \ldots, N,
\end{align*}
where
\begin{align*}
  \mathcal{H}_{\underline{\tilde{f}}}(0) &= \frac{1 - \sqrt{1 - 4h^{2}_{n}}}{2\pi 
    n h^{2}_{n}} \sum_{j = N - n + 1}^{N}\frac{1}{\lambda_{j}}, \\
  \tilde{\underline{f}}(\lambda_{i}) &= \frac{1}{n}\sum_{j=N - n + 1}^{N}
    \frac{\sqrt{[4\lambda^{2}_{j}h_{n}^{2} -
    (\lambda_{i}-\lambda_{j})^{2}]^{+}}}
    {2\pi \lambda^{2}_{j} h^{2}_{n}},\\
  \mathcal{H}_{\tilde{\underline{f}}}(\lambda_{i}) &= \frac{1}{n}
    \sum_{j= N - n + 1}^{N}\frac{\sgn(\lambda_{i} -
    \lambda_{j})\sqrt{[(\lambda_{i} - \lambda _{j})^{2} -
    4\lambda^{2}_{j}h_{n}^{2}]^{+}}  - \lambda_{i} + \lambda_{j}}{2\pi
    \lambda^{2}_{j} h^{2}_{n}}, 
\end{align*}
and  $h_{n} = n^{-\gamma}$ is the bandwidth with tuning
parameter~$\gamma$.

In both cases, the pool-adjacent-violators-algorithm (PAVA) in
isotonic regression can be used
to ensure the shrunken eigenvalues to be in ascending order. The
bandwidth parameter $\gamma$ can be selected via crossvalidation on
the estimated standard deviation of the scaling factors or
other information criteria. The feasible optimal nonlinear shrinkage
estimator is the resulting
$\hat{\Sigma}_{\mathrm{MV}} = \Gamma_n \tilde D_n  \Gamma_n^{\top} $.

\subsection{GTLS}\label{SubSec:gtls}
For the GTLS setting, which is more realistic with EIV, we propose to
pre-whiten $Y$ and $\tilde X_i$'s by $\hat\Sigma_{\mathrm{MV}}$ and
then apply the standard TLS procedure \citep{gleser1981estimation} to
estimate $\beta$. The resulting estimator of the $\beta$ will
be shown to be more efficient than that based on pre-whitening with
the linear shrinkage estimator $\hat\Sigma_{\mathrm{LS}}$
\citep{Ribe:Plan:Terr:appl:2013}.

Consider the GTLS estimator of $\beta$ obtained from prewhitening with
a class of regularized covariance matrix estimator $\hat\Sigma$ from
independent control runs. In the general framework of GTLS, the
measurement error vectors usually have the same covariance matrix as
the model error vector for the ease of theoretical derivations. This
assumption can be easily achived in the OF
setting~\eqref{eq:fp}--\eqref{eq:eiv} by multiplying
each observed fingerprint vector $\tilde X_i$ by $\sqrt{n_i}$.
Therefore, without loss of generality, in the
following we assume $n_i = 1$ to simplify the notations.

Let $\tilde X = (\tilde X_1, \ldots, \tilde X_p)$.  The GTLS estimator
based on $\hat\Sigma$ is
\begin{equation}
  \label{eq:obfun}
  \hat \beta(\hat\Sigma)
  =
  \underset{\beta}{\arg}\min \frac{\|\hat \Sigma^{-\frac{1}{2}} (Y -
    \tilde X \beta) \|^2_2}{1 + \beta^\top \beta},
\end{equation}
where $\|a\|_2$ is the $\ell_2$ norm of vector $a$.
The asymptotic properties of $\hat\beta(\hat\Sigma)$ are established
for a class of covariance matrix estimators $\hat\Sigma$ including
both $\hat\Sigma_{\mathrm{MV}}$ and $\hat\Sigma_{\mathrm{LS}}$.

\begin{assumption}
  \label{asmp:designmat}
  $\lim_{N, n \to \infty} X^\top \hat{\Sigma}^{-1} X / N = \Delta_1$
  exists, where $\Delta_1$ is a non-singular matrix.
\end{assumption}

\begin{assumption}
  \label{asmp:covmat}
  $\lim_{N, n \to \infty} \Tr(\hat{\Sigma}^{-1} \Sigma) / N$ exists and is a
  positive constant.
\end{assumption}

\begin{assumption}
  \label{asmp:eigen}
  $\lim_{N, n \to \infty} \Tr\{(\hat \Sigma^{-1/2} \Sigma \hat
  \Sigma^{-1/2})^2\} / N = K$ exists with $K > 0$.
\end{assumption}

\begin{remark}
  Assumptions~\ref{asmp:designmat} originates from
  \citet{gleser1981estimation}, which is needed for
  the consistency of $\hat\beta(\hat\Sigma)$.
  Assumptions~\ref{asmp:covmat}--\ref{asmp:eigen} are from
  \citet{ledoit2017optimal, ledoit2017direct}.
  Assumption~\ref{asmp:covmat} states that the average of the
  variances of the components of the pre-whitened error vectors
  converge to positive constant. For the class of rotation invariant
  estimators defined in \citet{ledoit2017nonlinear,
    ledoit2017direct}, which includes both $\hat\Sigma_{\mathrm{MV}}$
  and $\hat\Sigma_{\mathrm{LS}}$, Assumptions~\ref{asmp:covmat}
  and~\ref{asmp:eigen} are satisfied.
\end{remark}

\begin{lemma}
  \label{lem:consist}
  Under Assumptions~\ref{asmp:designmat}--\ref{asmp:eigen},
  $
  \hat \beta (\hat\Sigma) \overset{\mathcal{P}}{\to} \beta_0,
  $ as $N,n \to \infty$ with a $N / n \to c$ for some $c > 0$.
\end{lemma}

The proof for Lemma~\ref{lem:consist} is in Appendix~\ref{Sec:gtls}.

The asymptotic normality of $\hat \beta(\hat\Sigma)$ is established
with additional assumptions.

\begin{assumption}
  \label{asmp:weightedcrsp}
  $\lim_{N, n \to \infty} X^\top \hat \Sigma^{-1} \Sigma \hat \Sigma^{-1}
  X / N = \Delta_2$ exists for a non-singular matrix $\Delta_2$.
\end{assumption}

\begin{assumption}
  \label{asmp:normalerr}
  The regression error $\epsilon$ and measurement errors $\nu_i$'s,
  $i = 1, \ldots, p$, are mutually
  independent normally distributed random vectors.
\end{assumption}

\begin{remark}
  Assumption~\ref{asmp:weightedcrsp} originates from
  \citet{gleser1981estimation} for the asymptotic noramlity of the
  GTLS estimator.
  Assumption~\ref{asmp:normalerr} is commonly used in the context of
  climate change detection and attribution for mean state climate
  variables.
\end{remark}

\begin{theorem}
  \label{thm:asymnormal}
  Under Assumptions~\ref{asmp:designmat}--\ref{asmp:normalerr},
  as $N, n \to \infty$ with $N / n \to c$ for some $c > 0$,
  \begin{equation}
    \label{eq:asymvar}
    \sqrt N (\hat \beta - \beta_0)
    \overset{\mathcal{D}}{\to}
    \mathcal{N}(0, \Xi),
    \mbox{ where }
    \Xi = \Delta^{-1}_1
    \big\{\Delta_2 + K (I_p + \beta_0 \beta_0^\top)^{-1}\big\}
    (1 + \beta_0^\top\beta_0)\Delta^{-1}_1.
  \end{equation}
\end{theorem}

The proof of Theorem~\ref{thm:asymnormal} is in
Section~\ref{Sec:proof_asymn} of Appendix~\ref{Sec:gtls}.

The higher efficiency of the resulting estimator for $\beta$ from the
proposed weight matrix in comparison with that from the existing
weight is summarized by the following result with proof in
Section~\ref{Sec:proof_optimal} of Appendix~\ref{Sec:gtls}.

\begin{theorem}
  \label{prop:optimal}
  Let $\Xi(\hat\Sigma)$ be the asymptotic covariance matrix in
  Equation~\eqref{eq:asymvar} for a rotation invariant estimator
  $\hat\Sigma$ under
  Assumptions~\ref{asmp:designmat}--\ref{asmp:normalerr}.
  Then
  $\Tr\big( \Xi(\hat\Sigma_{\mathrm{MV}}) \big) \le
  \Tr\big( \Xi(\hat\Sigma_{\mathrm{LS}}) \big)$.
\end{theorem}



In our implementation, a 5-fold cross validation is used
to select the optimal bandwidth parameter $\gamma \in (0.2, 0.5)$.

\section{Simulation Studies}
\label{sec:numeric}

The finite sample performance of the proposed method in comparison
with the existing practice in RF needs to be assessed to make realistic
recommendations for detection and attribution of climate change. 
We conducted a simulation study similar to the setting of a study in
\citet{Ribe:Plan:Terr:appl:2013}. The observed climate variable of
interest is 11~decadal mean temperatures over 25~grid-boxes, a vector
of dimension $N = 275$.
Two $N\times 1$ fingerprints were considered, corresponding to the
anthropogenic (ANT) and natural forcings (NAT), denoted by $X_1$ and
$X_2$, respectively. They were set to the average of all runs from
the CNRM-CM5 model as in \cite{Ribe:Plan:Terr:appl:2013}.
To vary the strength of the signals, we also considered halving
$X_1$ and $X_2$. That is, there were two levels of signal-to-noise
ratio corresponding to the cases of multiplying each $X_i$,
$i \in \{1, 2\}$, controlled by a scale $\lambda \in \{1, 0.5\}$.
The true scaling factors were $\beta_1 = \beta_2 = 1$. The
distribution of the error vector $\epsilon$ was multivariate normal
$\mathrm{MVN}(0, \Sigma)$. The distribution of the
measurement error vector $\nu_i$, $i \in \{1, 2\}$, was
$\mathrm{MVN}(0, \Sigma/n_i)$, with $(n_1, n_2) = (35, 46)$ which are
the values in the detection and attribution analysis of annual mean
temperature conducted in Section~\ref{sec:real}.

Two settings of true $\Sigma$ were considered. In the first setting,
$\Sigma$ was an unstructured matrix $\Sigma_{\mathrm{UN}}$, which was
obtained by manipulating the eigenvalues but keeping the eigenvectors
of the proposed minimum variance estimate from the same set of climate
model simulations as in \cite{Ribe:Plan:Terr:appl:2013}. Specifically,
we first obtained the eigen decomposition of the minimum variance
estimate, and then restricted the eigenvalues to be equal over each of
the 25 grid-boxes (i.e., only 25 unique values for the $N = 25 \times
11$ eigenvalues) by taking averages over the decades at each grid-box.
The pattern of the resulting eigenvalues is similar to the pattern of
the eigenvalues of a spatial-temporal covariance matrix with
variance stationarity and weak dependence over the time dimension.
Finally, the eigenvalues were scaled independently by a uniformly
distributed variable on $[0.5, 1.5]$, which results in a more
unstructured covariance matrix setting similar to the simulation
settings in \citet{hannart2016integrated}.
In the second setting,  $\Sigma$ was set to be $\Sigma_{\mathrm{ST}}$
whose diagonals were set to be the sample variances from the climate
model simulations without imposing temporal stationarity; the
corresponding correlation matrix was set to be the Kronecker product
of a spatial correlation matrix and a temporal correlation matrix,
both with autoregressive of order~1 and coefficient~0.1.

The observed mean temperature vector $Y$ and
the estimated fingerprints $(\tilde X_1, \tilde X_2)$ were generated
from Models~\eqref{eq:fp}--\eqref{eq:eiv}. The control runs used to
estimate $\Sigma$ were generated from $\mathrm{MVN}(0, \Sigma)$ with
sample size $n \in \{50, 100, 200, 400\}$. For each replicate, the two
GTLS estimators of $\beta$ in Theorem~\ref{prop:optimal} were
obtained. For each configuration, 1000 replicates were run.

\begin{figure}[tbp]
  \centering
  \includegraphics[width=\textwidth]{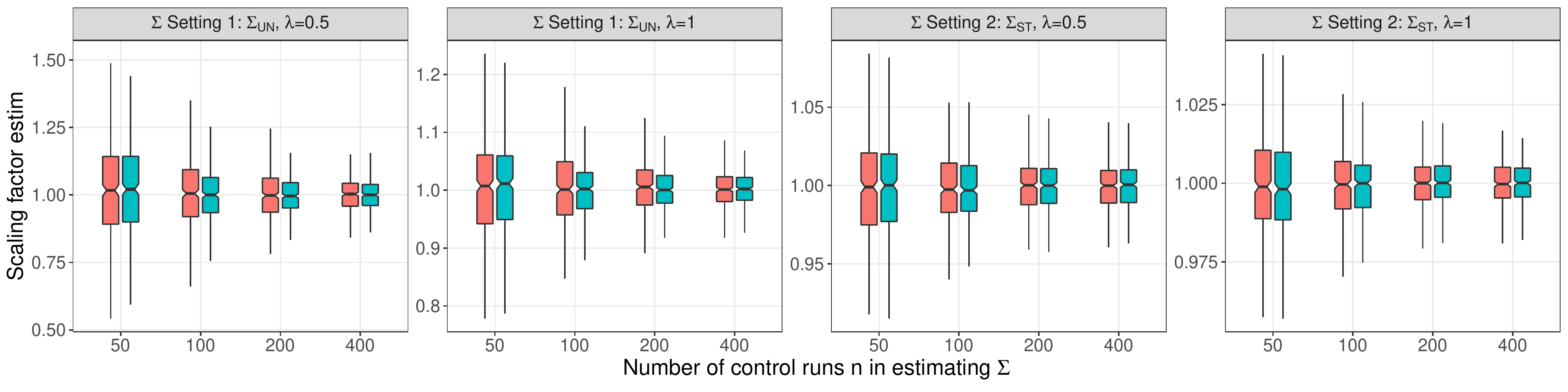}
  \caption{Boxplots of the estimates of the scaling factors in the
    simulation study based on 1000 replicates.}
  \label{fig:box}
\end{figure}


Figure~\ref{fig:box} displays the boxplots of the estimates of the ANT
scaling factor $\beta_1$ from the two TGLS approaches based on
pre-whitening with $\hat\Sigma_{\mathrm{LS}}$ (denoted as M1) and
$\hat\Sigma_{\mathrm{MV}}$ (denoted as M2) and, respectively.  Both
estimators appear to recover the true parameter values well on
average. The variations of both estimators are lower for larger $n$,
higher $\lambda$, and more structured $\Sigma$ (the case of
$\Sigma_{\mathrm{ST}}$). These observations are expected. A larger~$n$
means more accurate estimation $\Sigma$; a higher $\lambda$ means
stronger signal; a more structured $\Sigma$ means an easier task to
estimate $\Sigma$.  In the case of $\Sigma = \Sigma_{\mathrm{UN}}$,
the M2 estimates have smaller variations than the M1 estimate, since
the eigenvalues were less smooth and, hence, favored the nonlinear
shrinkage function.  For the case of $\Sigma = \Sigma_{ST}$ where the
covariance matrix is more structured, both methods estimate the true
covariance matrix much more accurately, and the differences between
methods are less obvious.  More detailed results are summarized in
Table~\ref{Tabap:existing_n_35_46} and
Table~\ref{Tabap:existing_n_10_6}, the latter of which had smaller
ensembles in estimating the fingerprints with $(n_1, n_2) = (10, 6)$.
The standard deviations of the M2 estimates are over 10\% smaller than
those of the M1 estimates for both cases.

\begin{figure}[tbp]
  \centering
  \includegraphics[width=\textwidth]{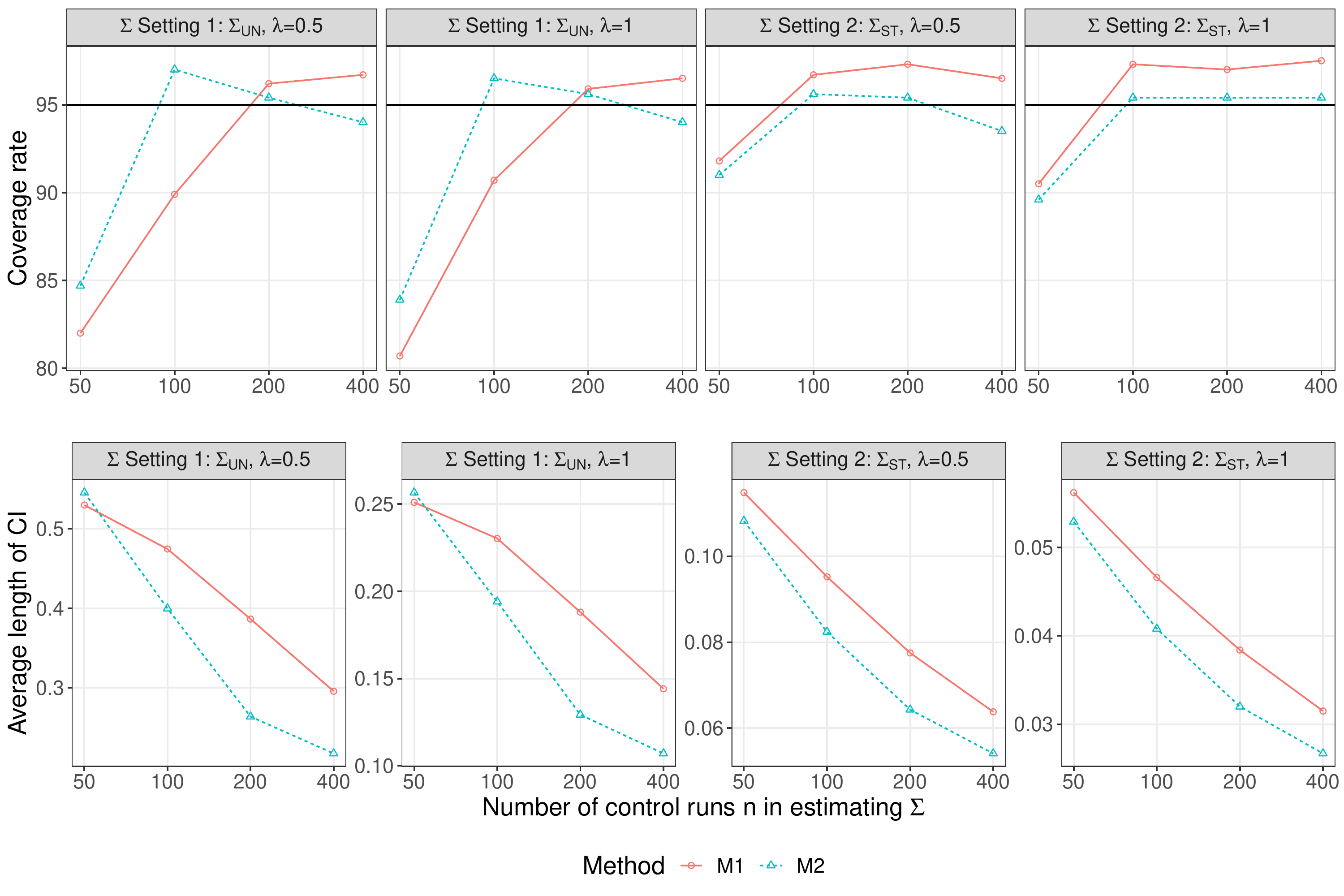}
  \caption{Calibrated 95\% confidence intervals for the scaling factors in the
    simulation study based on 1000 replicates.}
  \label{fig:ci}
\end{figure}

Confidence intervals are an important tool for detection and
attribution analyses. It would be desirable if the asymptotic variance
in Theorem~\ref{thm:asymnormal} can be used to construct confidence
intervals for the scaling factors. Unfortunately, it has been reported
that the confidence intervals for the scaling factors based on
$\hat\Sigma_{\mathrm{LS}}$ have coverage rates lower than, sometimes
much lower than, their nominal levels \citep{li2019confidence}. The
under-coverage issue remains for the estimator based on
$\hat\Sigma_{\mathrm{MV}}$. To give confidence intervals with correct
coverage rates, \citet{li2019confidence} proposed a calibration
procedure which enlarges the confidence intervals based on the
asymptotic normality of the estimators
by an appropriate scale tuned by a parametric bootstrap to
achieve the desired coverage rate. We applied this calibration to both
M1 and M2 estimators. Figure~\ref{fig:ci} shows the empirical coverage
rates of the 95\% confidence intervals after the calibration. The
coverage rate of a naive confidence interval could be as low as 70\%
(not shown). After the calibration, the coverage rates are much
closer to the nominal levels. The agreement is better for larger $n$
and more structured $\Sigma$. The calibrated confidence intervals from
M2 are about 10\% shorter to those from M1 overall in both settings of
true $\Sigma$, except for the case of $\Sigma = \Sigma_{\mathrm{UN}}$
and sample size $n=50$ where the confidence intervals suffer from
under-coverage issue..

\section{Fingerprinting Mean Temperature Changes}
\label{sec:real}

We apply the proposed approach to the detection and attribution
analyses of annual mean temperature of 1951--2010
at the global (GL), continental, and subcontinental scales. The
continental scale regions are
Northern Hemisphere (NH), NH midlatitude between
$30^\circ$ and $70^\circ$ (NHM), Eurasia (EA), and North America
(NA), which were studied in \citep{zhang2006multimodel}.
The subcontinental scale regions are Western North
American (WNA), Central North American (CNA), Eastern North
American (ENA), southern Canada (SCA), and southern Europe (SEU),
where the spatio-temporal correlation structure is more likely to hold.

In each regional analysis, we first constructed observation vector $Y$
from the HadCRUT4 dataset \citep{Colin2012HadCRUT4}.
The raw data were monthly anomalies of near-surface air
temperature on $5^\circ\times 5^\circ$ grid-boxes. At each grid-box,
each annual mean temperature was computed from monthly temperatures if
at least 9~months were available in that year; otherwise, it was
considered missing. Then, 5-year averages were computed if no more
than 2~annual averages were missing. To reduce the spatial dimension in
the analyses, the $5^\circ \times 5^\circ$ grid-boxes were aggregated
into larger grid-boxes. In particular, the grid-box sizes were
$40^\circ \times 30^\circ$ for GL and NH, $40^\circ \times 10^\circ$
for NH 30-70, $10^\circ \times 20^\circ$ for EA, and
$10^\circ \times 5^\circ$ for NA. For the subcontinent regions, no
aggregation was done except for SCA, in which case
$10^{\circ} \times 10^{\circ}$ grid-boxes were used. Details on the
longitude, latitude and spatio-temporal steps of each regions after
processing can be found in Table~\ref{tab:regions}.

\begin{table}[tbp]
  \centering
  \caption{\sf Summaries of the names, coordinate ranges, ideal
    spatio-temporal dimensions ($S$ and $T$'), and dimension of
    observation after removing missing values of the 5 regions
    analyzed in the study.}
  \label{tab:regions}
  \def\arraystretch{1}\tabcolsep=0.2em
  \begin{tabular}{llcccccc}
    \toprule
    Acronym & Regions & Longitude    & Latitude & Grid size & $S$ & $T$ & $n$ \\
                   &               &   ($^\circ$E) &   ($^\circ$N)   & ($1^\circ \times 1^\circ$) & & &\\
    \midrule
    \multicolumn{8}{c}{Global and Continental Regions}\\
    GL & Global & $-$180 / 180 & $-$90 / 90 & $40 \times 30$ & 54 & 11 & 572 \\
    NH & Northern Hemisphere & $-$180 / 180 & 0 / 90 & $40 \times 30$ & 27 & 11 & 297 \\
    NHM & Northern Hemisphere $30^\circ N$ to $70^\circ N$ & $-$180 / 180 & 30 / 70 & $40 \times 10$ & 36 & 11 & 396 \\
    EA & Eurasia & $-$10 / 180 & 30 / 70 & $10 \times 20$ & 38 & 11 & 418 \\
    NA & North America & $-$130 / $-$50 & 30 / 60 & $10 \times 5$ & 48 & 11 & 512 \\
    \addlinespace[1ex]
    \multicolumn{8}{c}{Subcontinental Regions}\\
    \midrule
    WNA & Western North America & $-$130 / $-$105 & 30 / 60 & $5 \times 5$ & 30 & 11 & 329 \\
    CNA & Central North America & $-$105 / $-$85 & 30 / 50 & $5 \times 5$ & 16 & 11 & 176 \\
    ENA & Eastern North America & $-$85 / $-$50 & 15 / 30 & $5 \times 5$ & 21 & 11 & 231 \\
    SCA & Southern Canada & $-$110 / $-$10 & 50 / 70 & $10 \times 10$ & 20 & 11 & 220 \\
    SEU & South Eupore & $-$10 / 40 & 35 / 50 & $5 \times 5$ & 30 & 11 & 330 \\
    \bottomrule
  \end{tabular}
\end{table}

Two external forcings, ANT and NAT, were considered.
Their fingerprints $X_1$ and $X_2$ were
not observed, but their estimates $\tilde X_1$ and $\tilde X_2$ were
averages over $n_1 = 35$ and $n_2 = 46$ runs from CIMP5 climate model
simulations. The missing pattern in $Y$ was used to
mask the simulated runs.
The same procedure used to aggregate the grid-boxes and obtain the
5-year averages in preparing $Y$ was applied to each masked run of each
climate model under each forcing. The final estimates $\tilde X_1$ and
$\tilde X_2$ at each grid-box were averages over all available runs
under the ANT and the NAT forcings, respectively, centered by the
average of the observed annual temperatures over 1961--1990, the same
center used by the HadCRUT4 data to obtain the observed anomalies.

Estimation of $\Sigma$ was based on $n = 223$ runs of 60~years
constructed from 47~pre-industrial control simulations of various
length. The long-term linear trend was removed separately from the
control simulations at each grid-box.  As the internal climate
variation is assumed to be stationary over time,
each control run was first split into
non-overlapping blocks of 60~years, and then each 60-year block was
masked by the same missing pattern as the HadCRUT4 data to create up
to 12 5-year averages at each grid-box. The temporal stationarity of
variance at each grid implies equal variance over time steps at each
observing grid-box, which is commonly incorporated in detection and
attribution analyses of climate change \citep[e.g.,][]{hannart2016integrated}.
Both M1 and M2 estimates based on linear and nonlinear shrinkage,
respectively, were obtained for comparison.
Pooled estimation of the variance at each grid-box was considered in
each of the shrinkage estimation to enforce the stationary, grid-box
specific variance.


\begin{figure}[tbp]
  \centering
  \includegraphics[width=\textwidth]{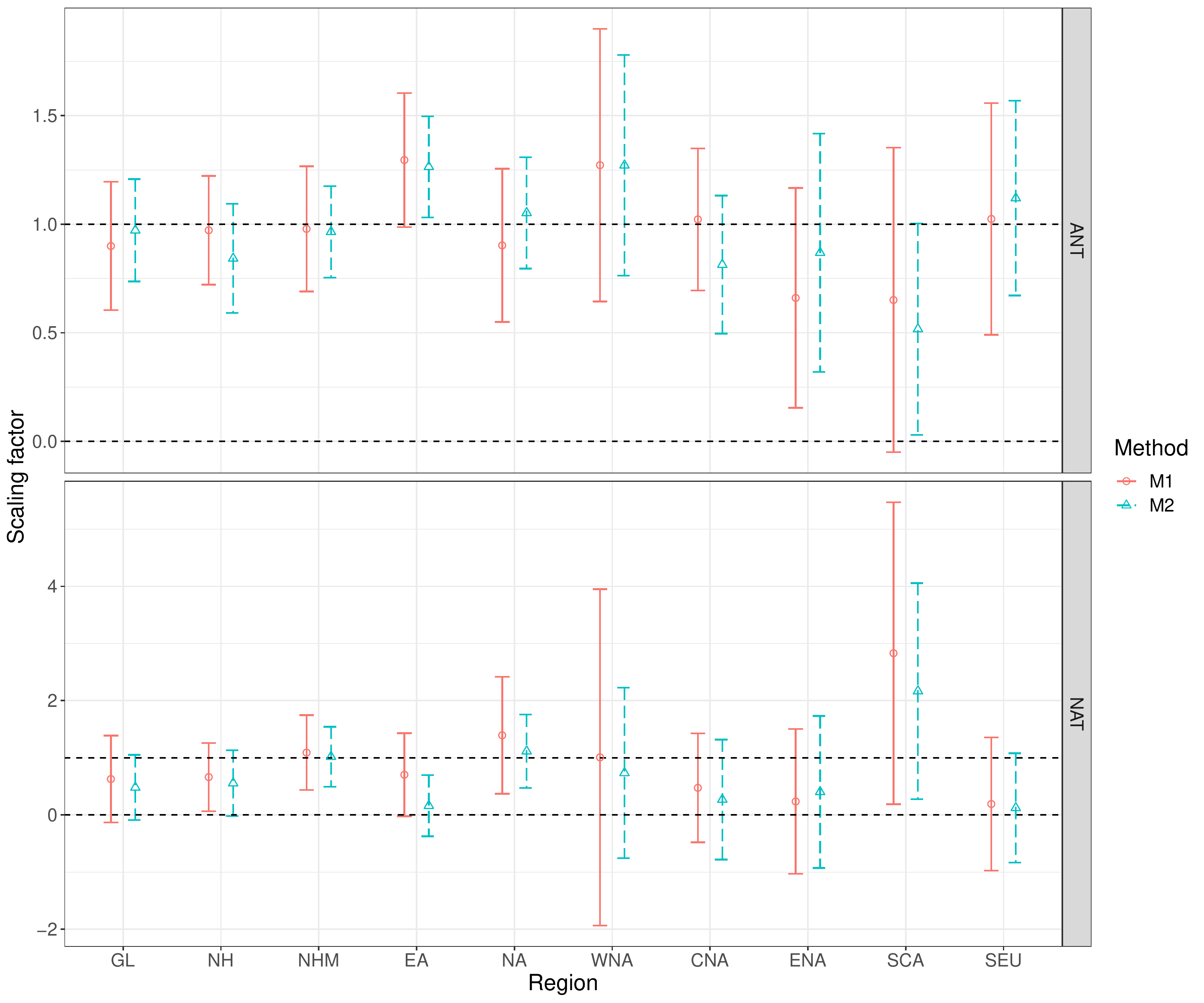}
  \caption{Estimated signal scaling factors for ANT and NAT
    required to best match observed 1950 - 2010 annual mean
    temperature for different spatial domains, and the corresponding
    95\% confidence intervals from different methods. For weight
    matrix construction, ``M1'' denotes the linear shrinkage approach
    and ``M2'' denotes the minimum variance approach. For confidence
    interval, the calibration method is used.}
  \label{fig:real_data}
\end{figure}

Figure~\ref{fig:real_data} summarizes the GTLS estimates of the
scaling factors $\hat\beta_1$ and $\hat\beta_2$ for the ANT and NAT
forcings, respectively. The estimates from pre-whitening weight matrix
$\Sigma_{\mathrm{LS}}$ and $\Sigma_{\mathrm{MV}}$ are denoted again as
M1 and M2, respectively. The 95\% confidence intervals were obtained
with the calibration approach of \citet{li2019confidence}. The point
estimates from M1 and M2 are similar in all the analysis.
The confidence intervals from the M2 method
are generally shorter than those from the M1 method in
the analyses both at continental and subcontinental
scale. More obvious reduction in the confidence interval
lengths is observed at the subcontinental scales, e.g., the ANT
scaling factor in EA/NA/SCA and the NAT scaling factor in NA/WNA/SCA.
This may be explained by that signals at subcontinental scale are
weaker and that the error covariance matrix has non-smooth
eigenvalues that form some clustering patterns due to weak temporal
dependence, as suggested by
the simulation study. Although the detection and attribution
conclusions based on the confidence intervals remain the same in most
cases, the shortened confidence intervals means reduced uncertainty in
the estimate of the attributable warming \citep{jones2013attribution}
and other quantities based on detection and attribution analysis, such
as future climate projection and transient climate sensitivity
\citep{li2019confidence}.

\section{Discussion}
\label{sec:disc}

Optimal fingerprinting as the most commonly used method for
detection and attribution analyses of climate change has great impact
in climate research. Such analyses are the basis for projecting
observationally constrained future climate
\citep[e.g.,][]{jones2016uncertainties} and
estimating important properties of the climate system such as climate
sensitivity \citep[e.g.,][]{schurer2018estimating}. The original
optimality of optimal fingerprinting, which minimizes the uncertainty in the
resulting scaling factor estimator, is no longer valid under realistic
settings where $\Sigma$ is not known but estimated.
Our method constructs a weight matrix by inverting a
nonlinear shrinkage estimator of $\Sigma$, which
directly minimizes the variation of the resulting scaling factor
estimator. This method is more efficient than the current RF practice
\citep{Ribe:Plan:Terr:appl:2013} as evident from the simulation study.
Therefore, the lost optimality in fingerprinting is restored
to a good extent for practical purposes, which helps to reduce the
uncertainty in important quantities such as attributable warming and
climate sensitivity.

There are open questions that we have not addressed.
It is of interest to further investigate how the asymptotic results
under $N, n \to \infty$ and $N/n \to c$ can guide the RF practice.
The temporal and spatial resolution that controls $N$ can be tuned in
RF practice, which may lead to different efficiencies in inferences
and, hence, different results in detection and attribution. Is there
an optimal temporal/spatial resolution to obtain the most reliable
result? Goodness-of-fit check is an important element in detection and
attribution analyses. The classic approach to check the weighted
sum of squared residuals against a chi-squared distribution under the
assumption of known $\Sigma$ is not valid when $\Sigma$ has to be
estimated. Can a test be designed, possibly based on parametric
bootstrap, to take into account of the uncertainty in regularized
estimation of $\Sigma$? These questions merit future research.


\section*{Acknowledgements}

 JY's research was partially supported by NSF grant DMS 1521730. KC's research was partially supported by NSF grant IIS 1718798.



\newpage
\setcounter{section}{0}

\renewcommand\thesection{\Alph{section}}
\renewcommand\theHsection{Appendix \Alph{section}}
\renewcommand\thesubsection{\Alph{section}.\arabic{subsection}}
\renewcommand\theHsubsection{\Alph{section}.\arabic{subsection}}
\renewcommand{\theequation}{\Alph{section}.\arabic{equation}}
\renewcommand{\theHequation}{\Alph{section}.\arabic{equation}}

\setcounter{table}{0}
\renewcommand{\thetable}{\Alph{section}.\arabic{table}}
\renewcommand{\theHtable}{\Alph{section}.\arabic{table}}
\setcounter{figure}{0}
\renewcommand{\thefigure}{\Alph{section}.\arabic{figure}}
\renewcommand{\theHfigure}{\Alph{section}.\arabic{figure}}

\begin{center}{\Large Supplementary Materials}\end{center}

\section{Sufficiency to Assume Orthogonal Covariates}
\label{Sec:orthlemma}

Consider the singular value decomposition of $N \times p$ design matrix $X$:
\[
X = UDV^{\top},
\]
where $U$ is a $N \times N$ orthogonal matrix, $D$ is a $N \times p$
matrix with $p$ non-negative singular values on the diagonal, and $V$ is
a $p \times p$ orthogonal matrix.

Let $X^* = UD = XV = (X_1^*, \ldots, X_p^*)$ and
$\beta^* = V^{\top}\beta = (\beta^{*}_1,\ldots, \beta^{*}_p)$.
The columns of $X^*$ are orthogonal.
The linear regression can be expressed as
\[
Y = \sum_{i=1}^p X^*_i \beta^*_i + \epsilon. 
\]

The estimator of $\beta^*$ is
\begin{align*}
  \hat{\beta}^*
  &= (X^{*^{\top}}\hat{\Sigma}^{-1}X^*)^{-1}X^{*^{\top}}\hat{\Sigma}^{-1}Y\\
  &= (V^{\top}X^{\top}\hat{\Sigma}^{-1}XV)^{-1}V^{\top}X^{\top}\hat{\Sigma}^{-1}Y\\
  &= V^{\top}(X^{\top}\hat{\Sigma}^{-1}X)^{-1}X^{\top}\hat{\Sigma}^{-1}Y
\end{align*}
and the corresponding covariance matrix is
\begin{align*}
  \Var(\hat{\beta}^*) = V^{\top}(X^{\top}\hat{\Sigma}^{-1}X)^{-1}
  \hat{\Sigma}^{-1}\Sigma \hat{\Sigma}^{-1}X(X^{\top}\hat{\Sigma}^{-1}X)^{-1}V.
\end{align*}
The orthogonality of $V$ ensures that
$\Tr(\Var(\hat{\beta})) = \Tr(\Var(\hat{\beta}^*)$ and
$\det(\Var(\hat{\beta})) = \det(\Var(\hat{\beta}^*)$.
Therefore, we only need to consider the orthogonal design in the
regression model.  In
other words, we may assume that the columns of the design matrix have
columns such that $X_i^{\top}X_j = 0$ for any $i \ne j$.

\section{Justification of Method M2 in the GLS Case}
\label{Sec:loss}

It is critically important to estimate the covariance matrix precisely as the
covariance matrix of $\hat\beta$ depends on $\hat\Sigma$.
In the estimation of the covariance matrix, an important question is how to
quantify the quality of the covariance matrix estimator. We use loss functions
to measure the quality of the covariance matrix estimator. For example, one loss
function is the Frobenius norm of the bias of the estimator
\[
  L(\hat{\Sigma}, \Sigma) = \|\hat{\Sigma}-\Sigma\|_F,
\]
and another loss function is Stein's loss
\[
L(\hat{\Sigma}, \Sigma) = \Tr(\hat{\Sigma}\Sigma^{-1}) -
\log \det(\hat{\Sigma}\Sigma^{-1}) - N,
\]
which is the Kullback--Leibler divergence under the normal assumption.

\subsection{Minimum Variance Loss Function}
\label{Sec:loss_trace_gls}

Considering the purpose of the fingerprinting, we construct a loss function
directly related to the variance of the scaling factor estimator. In other words,
we minimize the summation of the variances of the estimated scaling factors
\[
L_1(\hat{\Sigma}, \Sigma, X) =
\Tr((X^{\top}\hat{\Sigma}^{-1}X)^{-1}
X^{\top}\hat{\Sigma}^{-1}\Sigma \hat{\Sigma}^{-1}X
(X^{\top}\hat{\Sigma}^{-1}X)^{-1}).
\]

From the random matrix theory \citep{marcenko1967}, we have the
following results under the assumption that all limiting forms
exist. Consider an $N \times 1$ random vector
$\textbf{x} = (x_1, \ldots, x_N)^{^{\top}}$ whose entries are
independent and identically distributed with mean zero, positive
variance, and finite 12th moment.  Let $A$ be a given $N \times N$
matrix.  Then there exists a constant $K > 0$ independent of $N$ and
$\textbf{x}$ such that \citep{ledoit2011eigenvector,
  silverstein1995empirical}
\[
E|\textbf{x}^{\top}A\textbf{x} - \Tr(A)|^6 \le K\|A\|^6N^3,
\]
or
\[
E|\textbf{x}^{^{\top}}A\textbf{x}/N - \Tr(A)/N|^6 \le K\|A\|^6N^{-3},
\]
where $\|A\|$ is the spectral norm of $A$.
In other words,  if $\|A\| < C$ for some constant $C$, then as $ N\to \infty$,
\[
\textbf{x}^{^{\top}}A\textbf{x}/N - \Tr(A)/N \overset{a.s.}{\to} 0.
\]
Similarly, for two independent random vector $\textbf{x}$ and $\textbf{y}$,
\[
  \frac{\textbf{x}^{^{\top}}A\textbf{y}}{N} = \frac{(\textbf{x} +
    \textbf{y})^{^{\top}}A(\textbf{x} + \textbf{y})}{2N} -
  \frac{\textbf{x}^{^{\top}}A\textbf{x}}{2N} -
  \frac{\textbf{y}^{\top}A\textbf{y}}{2N}
  \overset{a.s.}{\to} 0.
\]

Now let
$X = (\textbf{x}_1, \textbf{x}_2, \ldots \textbf{x}_p)$
be an $N \times p$ matrix with independent columns.
Then, for any $N\times N$ matrix $A$ with $\|A\| < C$, the
$p \times p$ matrix
\[
\frac{X^{^{\top}}AX}{N} - \frac{\Tr(A)}{N}I_p \overset{a.s.}{\to} 0,
\]
where the righthand side is a $p\times p$ matrix of zeroes.
In other words, ${X^{^{\top}}AX}/{N}$ and ${\Tr(A)} I_p / N$ have the
same limit.

In optimal fingerprinting, let $\hat \Sigma_n$ be the sample
covariance matrix estimated from ensemble runs representing the
internal climate variation.  Suppose that the eigendecomposition of
$\hat \Sigma_n$ is $\hat \Sigma_n = \Gamma_n D_n \Gamma_n^{\top}$, where
$D_n = \diag(\lambda_1, \ldots, \lambda_N)$ is the diagonal matrix of the
eigenvalues and $\Gamma_n$ contains the corresponding eigenvectors.  Consider
the rotation invariant class of the estimators
$\hat{\Sigma} = \Gamma_n\tilde D_n \Gamma_n^{\top}$,
where $\tilde D_n = \diag(\delta(\lambda_i))$,
$i = 1, 2, \ldots, N$ for a smooth function $\delta(\cdot)$.  Under
the same Assumptions \ref{asmp:designmat}--\ref{asmp:eigen} of the main text, we
have both $\|\hat{\Sigma}^{-1}\|$ and
$\|\hat{\Sigma}^{-1}\Sigma \hat{\Sigma}^{-1}\|$ bounded.  Then the
assumptions of almost sure convergence are satisfied.

Consider the loss function
\begin{align*}
  \tilde{L}_1(\hat{\Sigma}, \Sigma, X)
  &= \frac{\Tr(X^{^{\top}}X)}{p^2}  L_1(\hat{\Sigma}, \Sigma, X)\\
  &= \frac{\Tr(X^{^{\top}}X)}{p^2}
    \Tr((X^{^{\top}}\hat{\Sigma}^{-1}X)^{-1}
    X^{^{\top}}\hat{\Sigma}^{-1}\Sigma \hat{\Sigma}^{-1}X
    (X^{^{\top}}\hat{\Sigma}^{-1}X)^{-1})\\
  &= \frac{\Tr(X^{^{\top}}X)}{p^2N}
    \Tr\left[\left(\frac{X^{^{\top}}\hat{\Sigma}^{-1}X}{N}\right)^{-1}
    \frac{X^{^{\top}}\hat{\Sigma}^{-1}\Sigma \hat{\Sigma}^{-1}X}{N}
    \left(\frac{X^{^{\top}}\hat{\Sigma}^{-1}X}{N}\right)^{-1}\right].
\end{align*}
Under the orthogonality assumption of $X$,
${\Tr(X^{\top}X)} / (pN)  \overset{a.s.}{\to} 1$.
Therefore, we can consider a loss function with the same limit,
\begin{align*}
L(\hat{\Sigma}, \Sigma) &= \frac{1}{p}
\Tr\left[ \left(\frac{\Tr(\hat{\Sigma}^{-1})}{N}I_p \right)^{-1}
\frac{\Tr(\hat{\Sigma}^{-1}\Sigma \hat{\Sigma}^{-1})}{N} I_p
\left(\frac{\Tr(\hat{\Sigma}^{-1})}{N}I_p\right)^{-1} \right]\\
&= \frac{\Tr(\hat{\Sigma}^{-1}\Sigma \hat{\Sigma}^{-1})/N}{(\Tr(\hat \Sigma^{-1}) / N)^2},
\end{align*}
which has the same form as what \citet{ledoit2017nonlinear} got.  We
have, as $ N \to \infty$,
\[
L(\hat{\Sigma}, \Sigma) - \tilde{L}_1(\hat{\Sigma}, \Sigma, X)
\overset{a.s.}{\to} 0.
\]

\subsection{Minimum General Variance Loss Function}

Instead of considering the trace, we can use the determinant of
$\Var(\hat\beta)$ as the objective loss function. Then the loss
function is
\begin{align*}
  L_2(\hat{\Sigma}, \Sigma, X)
  &= \det((X^{\top}\hat{\Sigma}^{-1}X)^{-1}X^{\top}\hat{\Sigma}^{-1}\Sigma \hat{\Sigma}^{-1}X
    (X^{\top}\hat{\Sigma}^{-1}X)^{-1})\\
  &= \frac{\det(X^{\top}\hat{\Sigma}^{-1}\Sigma \hat{\Sigma}^{-1}X)}{\det(X^{\top}\hat{\Sigma}^{-1}X)^2}.
\end{align*}
It is asymptotically equivalent to loss function
\begin{align*}
  \tilde{L}_2(\hat{\Sigma}, \Sigma, X)
  &= \left ( \frac{\Tr(X^{\top}X)}{p} \right)^p  L_2(\hat{\Sigma}, \Sigma, X)\\
  &= \left( \frac{\Tr(X^{\top}X)}{pN} \right)^p
    \frac{\det\left(\frac{X^{\top}\hat{\Sigma}^{-1}\Sigma
    \hat{\Sigma}^{-1}X}{N}\right)}
    {\det\left(\frac{X^{\top}\hat{\Sigma}^{-1}X}{N}\right)^2}.
\end{align*}

Similar to the minimum variance loss function, we can consider the
loss function with the same limiting form:
\begin{align*}
  L(\hat{\Sigma}, \Sigma)
 &= \det\left[ \left(\frac{\Tr(\hat{\Sigma}^{-1})}{N}I_p\right)^{-1}
   \frac{\Tr(\hat{\Sigma}^{-1}\Sigma \hat{\Sigma}^{-1})}{N} I_p
   \left(\frac{\Tr(\hat{\Sigma}^{-1})}{N} I_p \right)^{-1} \right]\\
 &= \left( \frac{\Tr(\hat{\Sigma}^{-1}\Sigma
   \hat{\Sigma}^{-1})/N}{(\Tr(\hat \Sigma^{-1}) / N)^2}\right)^p.
\end{align*}
In other words, as $N \to \infty$,
\[
L(\hat{\Sigma}, \Sigma) - \tilde{L}_2(\hat{\Sigma}, \Sigma, X)
\overset{a.s.}{\to} 0.
\]
That is, minimizing the limiting forms of two loss functions are
asymptotically equivalent to minimize the limiting form of
\[
L_{mv}(\hat{\Sigma}, \Sigma)
= \frac{\Tr(\hat{\Sigma}^{-1}\Sigma \hat{\Sigma}^{-1})/N}{(\Tr(\hat \Sigma^{-1}) / N)^2}.
\]
This concludes the proof of Theorem~\ref{thm:equal} of the main text.

\section{Justification of Method M2 in the GTLS Case}
\label{Sec:gtls}





\subsection{Proofs of Lemma~\ref{lem:consist} of the Main Text}
\label{Sec:proof_consist}

\begin{proof}
  The results are direct extensions of those in
  \citet{gleser1981estimation}, we only briefly sketch the
  proof. Consider the observed data matrix
  $\tilde A = \hat \Sigma^{-\frac{1}{2}} [\tilde X,  Y]$ which is
  obtained by binding the columns of $\tilde X$ and $Y$.
  Under the Assumption~\ref{asmp:designmat},~\ref{asmp:covmat} and
  \ref{asmp:eigen}, let $W = \tilde A^\top \tilde A$ and
  $\delta = \lim_{N \to \infty} \Tr(\hat{\Sigma}^{-1} \Sigma)/N$, we
  have
  \begin{equation} \label{eq:limcross}
    \lim_{N \to \infty} N^{-1} W
     = \delta I_{p + 1}  + (I_p, \beta_0)^\top \Delta_1 (I_p, \beta_0),
  \end{equation}
  which is in the same form of Lemma 3.1 in
  \citet{gleser1981estimation}. Consider the eigen decomposition of the
  matrix $W = G D G^\top$, where
  $D = \diag(d_1, \ldots, d_p, d_{p + 1})$, and the $(p + 1)^2$ matrix
  $G$ is partitioned as
  \[
    G =
    \begin{pmatrix}
      G_{11} & G_{12}\\
      G_{21} & G_{22}
    \end{pmatrix}
  \]
  with $p \times p$ matrix $G_{11}$.
  Then following the results of Lemma 3.3 by
  \citet{gleser1981estimation}, 
  $(G^{\top}_{11}, G^{\top}_{21})^{\top}$ converges to the
  eigenvectors corresponding to the $p$ largest eigenvalues of the
  limiting matrix of
  $\delta I_{p + 1} + (I_p, \beta_0)^\top \Delta_1 (I_p, \beta_0)$,
  denoted by $(H^{\top}_{11}, H^{\top}_{21})^{\top}$. Moreover, we
  have that there is a nonsingular $p\times p$ matrix $\psi$ such that
  \[
    \begin{pmatrix}
      H_{11} \\
      H_{21}
    \end{pmatrix} =
    \begin{pmatrix}
      I_p \\
      \beta^\top_{0}
    \end{pmatrix} \psi,
    \quad
    H_{21} H_{11}^{-1} = \beta^\top_{0}.
  \]

  The estimator from Equation~\eqref{eq:obfun} is given by
  $\hat \beta = \{G_{21}G^{-1}_{11}\}^{\top} = - G_{12} / G_{22}$.
  With the above results, we have
  \[
    \hat \beta \overset{\mathcal{P}}{\to} \beta_0,
  \]
  as $N, n \to \infty$ with $N / n \to c > 0$.
\end{proof}

\subsection{Proof of Theorem~\ref{thm:asymnormal} of the Main Text}
\label{Sec:proof_asymn}

\begin{proof}
  In the context of general asymptotics, i.e., $N, n \to \infty$ with fixed ratio,
  consider the eigendecomposition of matrix
  $\hat \Sigma^{-1/2} \Sigma \hat \Sigma^{-1/2} = U \Lambda U^\top$,
  where $\Lambda = \diag(\lambda_i)$ is the diagonal matrix of
  eigenvalues, and $U$ is the corresponding matrix of
  eigenvectors. Let
  \begin{align}
    \label{eq:transfer}
    \begin{split}
    &Y^* = U^\top \hat \Sigma^{-\frac{1}{2}} Y, \quad
      X^* = U^\top \hat \Sigma^{-\frac{1}{2}} X, \quad
      \epsilon^* =  U^\top \hat \Sigma^{-\frac{1}{2}} \epsilon, \\
    & \tilde X^* = U^\top \hat \Sigma^{-\frac{1}{2}} \tilde X, \quad
      V^* = U^\top \hat \Sigma^{-\frac{1}{2}} V.
    \end{split}
  \end{align}

  Then from Equation~\eqref{eq:obfun} of the main text, the generalized total
  least squares estimator $\hat \beta$ solves equation
  \[
    S(\beta)
    = \frac{\tilde X^{*\top}(y^* - \tilde X^* \beta)}{N}
    + \beta \frac{\|y^* - \tilde X^* \beta\|^2_F}
    {N(1 + \beta^\top \beta)}
    = 0,
  \]
  where $S(\beta) = (S_1(\beta), \ldots, S_p(\beta))^\top$ is a
  $p$ dimensional vector. By Taylor's theorem there exists a series of
  $\beta_j^*$ on the line segment between $\hat \beta$ and $\beta_0$
  for $j = 1, \ldots, p$ such that
  \begin{align*}
    & S_j(\hat \beta)
      = S_j(\beta_0) +
      [\nabla S_j(\beta_j^*)]^\top (\hat \beta - \beta_0) = 0,
      \quad j = 1, 2, ..., p \\
    & S (\hat \beta)
      = S(\beta_0) + H (\hat\beta - \beta_0) = 0,
      \quad H = (\nabla S_1(\beta_1^*), \ldots, \nabla S_p(\beta_p^*)).
  \end{align*}
  It follows that
  \begin{align*}
    H \sqrt{N} (\hat \beta - \beta_0)
    & = - \sqrt{N}S(\beta_0) \\
    & = - \frac{1}{\sqrt{N}} \sum_{i = 1}^N
      \left\{(\epsilon_i^* - v_i^{*\top} \beta_0) (x^*_i + v^*_i)
      + \beta_0 \frac{(\epsilon_i^* - v_i^{*\top} \beta_0)^2}
      {(1 + \beta_0^\top \beta_0)} \right\} \\
    & \coloneqq - \frac{1}{\sqrt{N}} \sum_{i = 1}^N f_i(\beta_0),
  \end{align*}
  where for $i = 1, \ldots, N$, $\epsilon^*_i$ is the $i$th element of
  vector $\epsilon$, $v^*_i$ is the $i$th row vector of $V^*$ and
  $x^*_i$ is the $i$th row vector of $X^*$.

  From Assumption~\ref{asmp:normalerr} in the main text and
  Equation~\eqref{eq:transfer}, 
  $\epsilon^*_i \sim N(0, \lambda_i)$,
  $v^*_i \sim \mathcal{N}(0, \lambda_i I_p)$, and $f_i(\beta_0)$ are mutually
  independent vectors with finite covariance matrices
  \begin{align*}
    \Var(f_i(\beta_0))
    & = \lambda_i x^*_i x^{*\top}_i (1 + \beta_0^\top \beta_0)
      + \lambda_i^2 (I_p + I_p \beta_0^\top \beta_0 + 2 \beta_0 \beta_0^\top)
      - 3 \lambda_i^2 \beta_0 \beta_0^\top \\
    & = [\lambda_i x^*_i x^{*\top}_i + \lambda_i^2 (I_p + \beta_0
      \beta_0^\top)^{-1}](1 + \beta_0^\top \beta_0),
  \end{align*}
  such that
  \[
    \lim_{N \to \infty} \frac{1}{N} \sum_{i = 1}^N \Var(f_i(\beta_0))
     = \big\{\Delta_2 + K (I_p + \beta_0 \beta_0^\top)^{-1} \big\}
      (1 + \beta_0^\top \beta_0).
  \]
  Then from Lemma~4.1 of \citet{gleser1981estimation} and
  Assumption~\ref{asmp:eigen} and~\ref{asmp:weightedcrsp} of the main text, 
  \begin{align*}
    \sqrt N S({\beta_0}) \overset{\mathcal{D}}{\to} MVN(0, \Xi_1),
  \end{align*}
  where $ \Xi_1 =
    \big\{\Delta_2 + K (I_p + \beta_0 \beta_0^\top)^{-1} \big\}
    (1 + \beta_0^\top \beta_0)$.

  It remains to show that as $N \to \infty$,
  \begin{equation} \label{eq:secondd}
    H = (\nabla S_1(\beta_1^*), \ldots, \nabla S_p(\beta_p^*))
    \overset{\mathcal{P}}{\to} \Delta_1.
  \end{equation}
  Consider the derivative of the score function $S(\beta)$ at
  any value of $\beta$
  \begin{align*}
    \nabla S(\beta)
    &= - \frac{\tilde X^{*\top} \tilde X^{*}}{N} +
      \frac{\|y^* - \tilde X^* \beta\|^2_F} {N(1 + \beta^\top
      \beta)} I_p + \beta \frac{\partial \|y^* - \tilde X^*
      \beta\|^2_F / \{N(1 + \beta^\top \beta)\}}{\partial \beta^\top}\\
    &= - \frac{\tilde X^{*\top} \tilde X^{*}}{N}
      + \frac{\|y^* - \tilde X^* \beta\|^2_F} {N(1 + \beta^\top \beta)} I_p
      + \frac{\beta \{S(\beta)\}^\top}{1 + \beta^\top \beta}.
  \end{align*}
  Since $S(\beta) \to 0$ as $N \to \infty$ and $\beta
  \overset{P}{\to} \beta_0$, we have
    \begin{align*}
      \lim_{N \to \infty, \beta \to \beta_0} \nabla S(\beta)
      &= \lim_{N \to \infty, \beta \to \beta_0} \left\{- \frac{\tilde X^{*\top} \tilde X^{*}}{N} +
        \frac{\|y^* - \tilde X^* \beta\|^2_F} {N(1 + \beta^\top \beta)} I_p \right\} \\
      & =  - \Delta_1
        -  \lim_{N \to \infty} \frac{\Tr(\hat \Sigma^{-1} \Sigma)}{N} I_p
        +  \lim_{N \to \infty} \frac{\Tr(\hat \Sigma^{-1} \Sigma)}{N} I_p \\
      & \phantom{=\,\,}
        + \lim_{N \to \infty, \beta \to \beta_0} \frac{(\beta - \beta_0)^\top \Delta_1 (\beta - \beta_0)}
        {(1 + \beta^\top \beta)} I_p \\
      & = - \Delta_1.
    \end{align*}
    Thus from the Lemma~\ref{lem:consist} of the main text and the fact that
    $\{\beta_j^*\}$ is a sequence on the line segment between
    $\hat \beta$ and $\beta_0$ for $j = 1, \ldots, p$,
    Equation~\eqref{eq:secondd} holds. With this, we complete the proof
    of Theorem~\ref{thm:asymnormal}.
\end{proof}

\subsection{Proof of Theorem~\ref{prop:optimal} of the Main Text}
\label{Sec:proof_optimal}

\begin{proof}
  Here we sketch the proof in the context of optimal
  fingerprinting. More rigorous derivations are to be established for
  the more general setting. As mentioned
  in Section~\ref{SubSec:gtls} of the main text, without
  loss of generality, we adjust the Model~\eqref{eq:eiv} to make
  the covariance matrix of the errors in the covariartes and in the
  response the same. Consider model
  \begin{align*}
    Y &= \sum_{i=1}^p X^*_i \beta^*_i + \epsilon,\\
    \tilde X^*_i &= X^*_i + \nu^*_i, \qquad i = 1, \ldots, p,
  \end{align*}
  where $\nu^*_i$ is a normally distributed measurement error vector
  with mean zero and covariance matrix $\Sigma/n_i$. Here for the ease
  of illustration, we consider $n_1= \ldots = n_p = n_0$. Usually in the
  above model, the magnitudes of fingerprints are comparable to the
  noise in the sense that the values of $\Tr\{(X^*)^\top X^*\}$ and
  $\Tr(\Sigma)$ are comparable, i.e.,
  $\Tr\{(X^*)^\top X^*\}/p \Tr(\Sigma) = O(1)$ as $N \to \infty$.

  Let $\tilde X_i = \sqrt{n_0} \tilde X^*_i$,
  $X_i = \sqrt{n_0} X^*_i$, $\nu_i = \sqrt{n_0} \nu^*_i$ and
  $\beta_i = \beta^*_i / \sqrt{n_0}$. Then the model can be rewritten
  as
  \begin{align} \label{eq:eiv2}    
    Y &= \sum_{i=1}^p X_i \beta_i + \epsilon,\\
    \tilde X_i &= X_i + \nu_i, \qquad i = 1, \ldots, p,
  \end{align}
  which is exactly in the form of Models~\eqref{eq:fp}--\eqref{eq:eiv}
  of the main text with $n_i = 1$. Then the theoretical results in
  Section~\ref{SubSec:gtls} of the main text are directly applied to
  the adjusted model. The coefficient estimations and corresponding
  variance estimations of the original model can be easily obtained
  from the results of above adjusted model. That is, in the context of
  optimal fingerprinting, the original model is equivalent to fit
  models with relative large magnitudes of fingerprints $X_i$ and
  small values of true coefficients $\beta_0$.

  Consider the trace of asymptotic covariance matrix for the estimated
  coefficient vector in Model~$\eqref{eq:eiv2}$ given by
  \begin{align}\label{eq:trace}
    \Tr(\Xi)
    &= \Tr(\Delta^{-1}_1 \big\{\Delta_2 + K (I_p + \beta_0
      \beta_0^\top)^{-1}\big\} (1 + \beta_0^\top\beta_0)\Delta^{-1}_1)\\
    &= \Tr(\Delta^{-1}_1 \Delta_2 \Delta^{-1}_1) + (1 +
      \beta_0^\top\beta_0) \Tr(\Delta^{-1}_1 K (I_p + \beta_0
      \beta_0^\top)^{-1}\Delta^{-1}_1)
  \end{align}
  from the Theorem~\ref{thm:asymnormal} of the main text.

  The first term $\Tr(\Delta^{-1}_1 \Delta_2 \Delta^{-1}_1)$ on
  the right hand side of \eqref{eq:trace} is the same as the loss
  function proposed in Appendix~\ref{Sec:loss_trace_gls}. In the
  context of general asymptotics, i.e., $N, n \to \infty$ with fixed
  ratio, the first term is equivalent to
  \[
    \frac{Np^2}{\Tr(X^\top X)} \frac{\Tr(\hat{\Sigma}^{-1}\Sigma
      \hat{\Sigma}^{-1})/N}{(\Tr(\hat \Sigma^{-1}) / N)^2},
  \]
  as $X^\top A X / N$ is asymptotically equivalent to
  $ \{\Tr(A) I_p / N\}\{\Tr(X^\top X) / Np\}$ for properly defined
  matrices $X$ and $A$ (See Appendix~\ref{Sec:loss_trace_gls}).

  As for the second term
  $(1 + \beta_0^\top\beta_0) \Tr(\Delta^{-1}_1 K (I_p + \beta_0
  \beta_0^\top)^{-1}\Delta^{-1}_1)$, we need to show that it is
  dominated by the first term. Given the information on the magnitudes
  of fingerprints $X$ and coefficients $\beta_0$ from the above
  illustrations, with appropriate large choice for the value of $n_0$
  (which is fairly reasonable in real fingerprinting studies), the
  small values of $\beta_0$ can be omitted in the second term. Then
  the second term can be approximated by
  $\Tr(\Delta^{-1}_1 K \Delta^{-1}_1)$. We further note that
  $K = \lim_{N, n \to \infty} \Tr\{(\hat \Sigma^{-1/2} \Sigma \hat
  \Sigma^{-1/2})^2\} / N$ and
  $\Tr\{(\hat \Sigma^{-1/2} \Sigma \hat \Sigma^{-1/2})^2\} / N \le
  \{\Tr(\hat \Sigma^{-1} \Sigma \hat \Sigma^{-1})\Tr(\Sigma)\} /
  N$. Then we have in the general asymptotics
  \begin{align*}
    \Tr(\Delta^{-1}_1 K \Delta^{-1}_1)
    &\le \frac{N^2p^3}{\Tr(X^\top X)^2} \frac{\{\Tr(\hat \Sigma^{-1} \Sigma \hat
      \Sigma^{-1})\Tr(\Sigma)\}/N}{(\Tr(\hat \Sigma^{-1}) / N)^2}\\
    &= \frac{Np^2}{\Tr(X^\top X)} \frac{\{\Tr(\hat \Sigma^{-1} \Sigma \hat
      \Sigma^{-1})\}/N}{(\Tr(\hat \Sigma^{-1}) / N)^2} \frac{\Tr(\Sigma)}{\Tr(X^\top X)/p},
  \end{align*}
  where $\Tr(\Sigma)/(\Tr(X^\top X)/p) = O(1/n_0)$ as $N \to \infty$
  based on the adjustment in Model~\eqref{eq:eiv2}. That is, with
  fairly large value of the number of climate simulations to obtain
  the estimated fingerprints $\tilde X_i$, the second term on the
  right hand side of \eqref{eq:trace} can be dominated by the first
  term, i.e., as $N \to \infty$ we have
  \begin{align*}
    (1 + \beta_0^\top\beta_0) \Tr(\Delta^{-1}_1 K (I_p + \beta_0
    \beta_0^\top)^{-1}\Delta^{-1}_1)
    &\le \delta \Tr(\Delta^{-1}_1 \Delta_2 \Delta^{-1}_1),
  \end{align*}
  for an arbitrary small value $\delta$ with large enough magnitude of
  $X_i$ in Model~\eqref{eq:eiv2}, i.e., large enough number of climate
  simulations for computing the fingerprints in the original model.
  Thus the proposed covariance matrix $\hat\Sigma_{\mathrm{MV}}$ which
  is the optimal choice regarding the first term in general
  asymptotics is expected to outperform the linear shrinkage estimator
  $\hat\Sigma_{\mathrm{LS}}$ in the sense that
  \[
    \Tr\big( \Xi(\hat\Sigma_{\mathrm{MV}}) \big) \le \Tr\big(
    \Xi(\hat\Sigma_{\mathrm{LS}}) \big),
  \]
  which is consistent with the fact that as the number of climate
  simulations to estimate the fingerprints becomes larger, the effects
  of measurement error diminish.

  This completes the proof.
\end{proof}

\section{Detailed Results on Simulation Studies}
\label{Sec:results}

\begin{figure}[tbp]
  \centering
  \includegraphics[width=\textwidth]{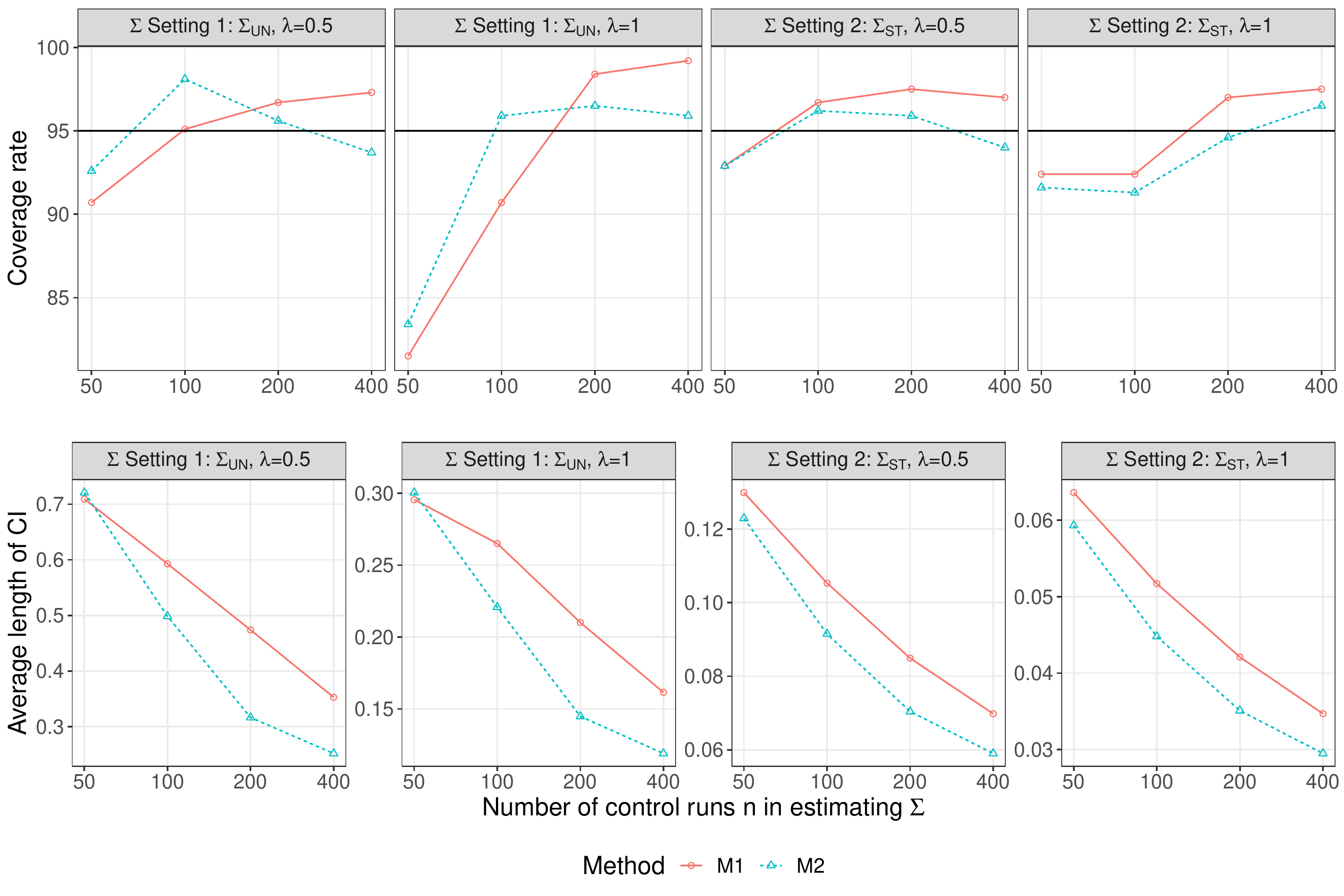}
  \caption{Calibrated 95\% confidence intervals for the scaling
    factors in the simulation study based on 1000 replicates. The
    numbers of runs for two forcings are $n_1=10$ and $n_2 = 6$
    respectively.}
  \label{fig:ci_10_6}
\end{figure}

The results of simulation studies in Section~\ref{sec:numeric} of the
main text are detailed in Table~\ref{Tabap:existing_n_35_46}, Table~\ref{Tabap:existing_n_10_6} and Figure~\ref{fig:ci_10_6}.


\begin{table}
  \centering
  \caption{Scaling factor estimates and 95\% confidence interval. Two
    structures of covariance matrix is considered:
    $\Sigma_{\mathrm{UN}}$ for proposed shrinkage estimator from
    ensemble simulations and $\Sigma_{\mathrm{ST}}$ for separable
    spatio-temporal covariance matrix. $\lambda$ is the scale to
    control the signal-to-noise ratio. Error terms are generated from
    multivariate normal distribution. Number of ensembles for two
    forcings are $n_1=35$ and $n_2=46$. Two constructions of weight
    matrix are compared. ``M1'' denote for linear shrinkage method and
    ``M2'' denote for proposed approaches. Existing formula-based
    method (N) and calibration method (CB) are used to construct 95\%
    confidence intervals. Bias and standard deviation (SD $\times$
    100) of scaling factors and average length of confidence intervals
    (CIL) and corresponding empirical rate (CR) from 1000 replicates
    are recorded. } \label{Tabap:existing_n_35_46} \footnotesize
  \begin{tabular}{llrrrrrrrrrrrr}
  \toprule
    & &  \multicolumn{6}{c}{ANT} & \multicolumn{6}{c}{NAT}\\
  \cmidrule(lr){3-8} \cmidrule(lr){9-14}
    & & & &
  \multicolumn{2}{c}{N} & \multicolumn{2}{c}{CB} & & &
  \multicolumn{2}{c}{N} & \multicolumn{2}{c}{CB} \\
  \cmidrule(lr){5-6} \cmidrule(lr){7-8} \cmidrule(lr){11-12}
  \cmidrule(lr){13-14}
  size & method & Bias & SD & CIL & CR & CIL & CR & Bias & SD & CIL & CR & CIL & CR \\
  \midrule
  \multicolumn{14}{c}{$\Sigma$ Setting 1: $\Sigma_{\mathrm{UN}}$; SNR $\lambda = 0.5$}\\
    50 & M1 & 0.02 & 19.0 & 0.41 & 69.8 & 0.53 & 82.0 & 0.01 & 57.0 & 1.47 & 78.5 & 1.89 & 89.9 \\
    & M2 & 0.02 & 17.3 & 0.38 & 69.2 & 0.55 & 84.7 & 0.01 & 49.7 & 1.23 & 76.3 & 1.79 & 92.6 \\
    100 & M1 & 0.01 & 13.6 & 0.36 & 76.8 & 0.47 & 89.9 & -0.02 & 36.9 & 1.12 & 85.8 & 1.46 & 94.0 \\
    & M2 & -0.00 & 9.7 & 0.29 & 83.4 & 0.40 & 97.0 & -0.02 & 26.8 & 0.81 & 86.1 & 1.13 & 95.6 \\
    200 & M1 & 0.00 & 9.0 & 0.30 & 89.4 & 0.39 & 96.2 & -0.01 & 24.4 & 0.87 & 92.4 & 1.08 & 97.5 \\
    & M2 & -0.00 & 6.7 & 0.23 & 91.3 & 0.26 & 95.4 & -0.01 & 18.7 & 0.64 & 89.1 & 0.76 & 94.8 \\
    400 & M1 & 0.00 & 6.1 & 0.26 & 93.7 & 0.30 & 96.7 & -0.01 & 18.9 & 0.72 & 93.2 & 0.83 & 97.0 \\
    & M2 & -0.00 & 5.6 & 0.20 & 93.2 & 0.22 & 94.0 & -0.00 & 16.1 & 0.57 & 92.1 & 0.64 & 94.3 \\
    \midrule
    \multicolumn{14}{c}{$\Sigma$ Setting 1: $\Sigma_{\mathrm{UN}}$; SNR $\lambda = 1$}\\
    50 & M1 & 0.00 & 9.0 & 0.20 & 71.9 & 0.25 & 80.7 & -0.01 & 20.4 & 0.63 & 87.2 & 0.73 & 92.4 \\
    & M2 & 0.00 & 8.5 & 0.18 & 72.5 & 0.26 & 83.9 & -0.00 & 19.2 & 0.55 & 80.9 & 0.68 & 89.6 \\
    100 & M1 & 0.00 & 6.6 & 0.18 & 79.8 & 0.23 & 90.7 & -0.01 & 14.8 & 0.51 & 91.0 & 0.59 & 96.2 \\
    & M2 & 0.00 & 4.5 & 0.14 & 85.6 & 0.19 & 96.5 & -0.01 & 12.0 & 0.38 & 86.9 & 0.46 & 94.6 \\
    200 & M1 & 0.00 & 4.5 & 0.15 & 88.6 & 0.19 & 95.9 & -0.00 & 10.2 & 0.41 & 95.4 & 0.46 & 97.5 \\
    & M2 & 0.00 & 3.3 & 0.12 & 92.1 & 0.13 & 95.6 & 0.00 & 8.7 & 0.31 & 93.2 & 0.35 & 95.9 \\
    400 & M1 & 0.00 & 3.2 & 0.13 & 94.0 & 0.14 & 96.5 & -0.00 & 8.3 & 0.34 & 95.6 & 0.37 & 96.7 \\
    & M2 & 0.00 & 2.8 & 0.10 & 92.4 & 0.11 & 94.0 & -0.00 & 7.8 & 0.28 & 91.3 & 0.30 & 95.4 \\
    \midrule
    \multicolumn{14}{c}{$\Sigma$ Setting 2: $\Sigma_{\mathrm{ST}}$; SNR $\lambda = 0.5$}\\
    50 & M1 & -0.00 & 3.2 & 0.10 & 87.7 & 0.11 & 91.8 & 0.00 & 8.7 & 0.33 & 91.8 & 0.39 & 96.7 \\
    & M2 & 0.00 & 3.1 & 0.09 & 85.6 & 0.11 & 91.0 & 0.00 & 8.5 & 0.30 & 89.9 & 0.35 & 95.1 \\
    100 & M1 & -0.00 & 2.2 & 0.08 & 93.2 & 0.10 & 96.7 & 0.00 & 7.1 & 0.26 & 91.3 & 0.30 & 95.6 \\
    & M2 & -0.00 & 2.0 & 0.07 & 91.6 & 0.08 & 95.6 & 0.00 & 7.0 & 0.23 & 87.7 & 0.26 & 91.6 \\
    200 & M1 & -0.00 & 1.6 & 0.07 & 95.6 & 0.08 & 97.3 & 0.00 & 5.9 & 0.22 & 94.0 & 0.25 & 96.5 \\
    & M2 & -0.00 & 1.6 & 0.06 & 94.0 & 0.06 & 95.4 & 0.00 & 5.6 & 0.19 & 90.5 & 0.21 & 93.7 \\
    400 & M1 & -0.00 & 1.5 & 0.06 & 94.8 & 0.06 & 96.5 & 0.00 & 5.1 & 0.19 & 91.0 & 0.21 & 93.7 \\
    & M2 & -0.00 & 1.4 & 0.05 & 92.4 & 0.05 & 93.5 & -0.00 & 4.8 & 0.17 & 88.8 & 0.18 & 90.7 \\
    \midrule
    \multicolumn{14}{c}{$\Sigma$ Setting 2: $\Sigma_{\mathrm{ST}}$; SNR $\lambda = 1$}\\
    50 & M1 & -0.00 & 1.6 & 0.05 & 84.2 & 0.06 & 90.5 & 0.00 & 4.4 & 0.16 & 93.5 & 0.18 & 94.8 \\
    & M2 & -0.00 & 1.6 & 0.04 & 80.7 & 0.05 & 89.6 & 0.00 & 4.5 & 0.15 & 90.2 & 0.16 & 91.6 \\
    100 & M1 & -0.00 & 1.1 & 0.04 & 92.4 & 0.05 & 97.3 & 0.00 & 3.2 & 0.13 & 95.4 & 0.14 & 97.8 \\
    & M2 & -0.00 & 1.0 & 0.04 & 89.4 & 0.04 & 95.4 & 0.00 & 3.2 & 0.11 & 92.6 & 0.12 & 95.9 \\
    200 & M1 & -0.00 & 0.8 & 0.03 & 94.0 & 0.04 & 97.0 & 0.00 & 2.8 & 0.11 & 94.8 & 0.12 & 96.7 \\
    & M2 & 0.00 & 0.7 & 0.03 & 93.7 & 0.03 & 95.4 & 0.00 & 2.7 & 0.10 & 92.6 & 0.10 & 94.6 \\
    400 & M1 & 0.00 & 0.7 & 0.03 & 96.7 & 0.03 & 97.5 & 0.00 & 2.3 & 0.09 & 95.9 & 0.10 & 97.8 \\
    & M2 & 0.00 & 0.6 & 0.03 & 95.4 & 0.03 & 95.4 & 0.00 & 2.3 & 0.08 & 92.4 & 0.09 & 94.6 \\
  \bottomrule
\end{tabular}
\end{table}

\begin{table}
  \centering
  \caption{Scaling factor estimates and 95\% confidence interval. Two
    structures of covariance matrix is considered:
    $\Sigma_{\mathrm{UN}}$ for proposed shrinkage estimator from
    ensemble simulations and $\Sigma_{\mathrm{ST}}$ for separable
    spatio-temporal covariance matrix. $\lambda$ is the scale to
    control the signal-to-noise ratio. Error terms are generated from
    multivariate normal distribution. Number of ensembles for two
    forcings are $n_1=10$ and $n_2=6$. Two constructions of weight
    matrix are compared. ``M1'' denote for linear shrinkage estimator
    and ``M2'' denote for proposed approach. Existing formula-based
    method (N) and calibration method (CB) are used to construct 95\%
    confidence intervals. Bias and standard deviation (SD $\times$
    100) of scaling factors and average length of confidence intervals
    (CIL) and corresponding empirical rate (CR) from 1000 replicates
    are recorded.} \label{Tabap:existing_n_10_6} \footnotesize
  \begin{tabular}{llrrrrrrrrrrrr}
  \toprule
    & &  \multicolumn{6}{c}{ANT} & \multicolumn{6}{c}{NAT}\\
  \cmidrule(lr){3-8} \cmidrule(lr){9-14}
    & & & &
  \multicolumn{2}{c}{N} & \multicolumn{2}{c}{CB} & & &
  \multicolumn{2}{c}{N} & \multicolumn{2}{c}{CB} \\
  \cmidrule(lr){5-6} \cmidrule(lr){7-8} \cmidrule(lr){11-12}
  \cmidrule(lr){13-14}
  size & method & Bias & SD & CIL & CR & CIL & CR & Bias & SD & CIL & CR & CIL & CR \\
  \midrule
    \multicolumn{14}{c}{$\Sigma$ Setting 1: $\Sigma_{\mathrm{UN}}$; SNR $\lambda = 0.5$}\\
    50 & M1 & 0.00 & 24.6 & 0.53 & 74.9 & 0.71 & 90.7 & -0.00 & 130.9 & 2.08 & 66.8 & 3.07 & 85.8 \\
    & M2 & 0.00 & 22.8 & 0.47 & 71.9 & 0.72 & 92.6 & 0.05 & 115.2 & 1.77 & 64.9 & 3.22 & 92.1 \\
    100 & M1 & -0.00 & 17.4 & 0.44 & 82.3 & 0.59 & 95.1 & 0.06 & 97.1 & 1.90 & 70.0 & 2.90 & 90.7 \\
    & M2 & -0.00 & 11.5 & 0.34 & 86.4 & 0.50 & 98.1 & 0.02 & 53.3 & 1.29 & 76.6 & 2.28 & 94.6 \\
    200 & M1 & 0.01 & 11.4 & 0.36 & 89.1 & 0.47 & 96.7 & 0.03 & 47.0 & 1.37 & 83.7 & 2.02 & 94.6 \\
    & M2 & 0.00 & 7.4 & 0.27 & 92.1 & 0.32 & 95.6 & 0.00 & 29.9 & 0.91 & 88.8 & 1.24 & 97.5 \\
    400 & M1 & 0.00 & 7.5 & 0.30 & 94.3 & 0.35 & 97.3 & 0.02 & 30.5 & 1.09 & 90.5 & 1.41 & 96.2 \\
    & M2 & 0.00 & 6.5 & 0.23 & 91.8 & 0.25 & 93.7 & 0.01 & 21.3 & 0.79 & 92.4 & 0.95 & 96.2 \\
  \midrule
  \multicolumn{14}{c}{$\Sigma$ Setting 1: $\Sigma_{\mathrm{UN}}$; SNR $\lambda = 1$}\\
    50 & M1 & -0.00 & 11.0 & 0.23 & 69.2 & 0.30 & 81.5 & 0.05 & 37.7 & 0.90 & 76.8 & 1.22 & 87.7 \\
    & M2 & -0.01 & 10.5 & 0.21 & 69.2 & 0.30 & 83.4 & 0.03 & 32.7 & 0.73 & 72.5 & 1.15 & 92.4 \\
    100 & M1 & -0.00 & 7.6 & 0.20 & 78.5 & 0.27 & 90.7 & 0.01 & 24.6 & 0.69 & 82.3 & 0.94 & 92.9 \\
    & M2 & 0.00 & 5.2 & 0.16 & 86.9 & 0.22 & 95.9 & 0.00 & 17.0 & 0.47 & 83.4 & 0.71 & 95.1 \\
    200 & M1 & -0.00 & 4.7 & 0.17 & 91.0 & 0.21 & 98.4 & -0.00 & 15.0 & 0.50 & 88.8 & 0.66 & 95.4 \\
    & M2 & -0.00 & 3.4 & 0.13 & 94.6 & 0.14 & 96.5 & -0.01 & 11.5 & 0.36 & 87.2 & 0.45 & 94.0 \\
    400 & M1 & 0.00 & 3.4 & 0.14 & 97.8 & 0.16 & 99.2 & -0.00 & 11.4 & 0.42 & 90.2 & 0.49 & 95.4 \\
    & M2 & 0.00 & 2.8 & 0.11 & 95.1 & 0.12 & 95.9 & -0.01 & 9.3 & 0.32 & 89.6 & 0.37 & 92.4 \\
  \midrule
  \multicolumn{14}{c}{$\Sigma$ Setting 2: $\Sigma_{\mathrm{ST}}$; SNR $\lambda = 0.5$}\\
    50 & M1 & 0.00 & 3.6 & 0.11 & 85.6 & 0.13 & 92.9 & 0.01 & 14.2 & 0.39 & 81.2 & 0.60 & 96.2 \\
    & M2 & 0.00 & 3.5 & 0.10 & 83.4 & 0.12 & 92.9 & 0.01 & 12.6 & 0.35 & 81.5 & 0.55 & 97.3 \\
    100 & M1 & -0.00 & 2.5 & 0.09 & 91.8 & 0.11 & 96.7 & -0.00 & 9.0 & 0.30 & 88.0 & 0.40 & 96.7 \\
    & M2 & -0.00 & 2.3 & 0.08 & 91.6 & 0.09 & 96.2 & -0.00 & 8.1 & 0.27 & 89.1 & 0.34 & 95.1 \\
    200 & M1 & 0.00 & 1.9 & 0.07 & 94.6 & 0.08 & 97.5 & 0.00 & 6.6 & 0.25 & 93.5 & 0.30 & 97.3 \\
    & M2 & 0.00 & 1.8 & 0.06 & 94.3 & 0.07 & 95.9 & 0.00 & 5.9 & 0.22 & 92.6 & 0.25 & 96.5 \\
    400 & M1 & -0.00 & 1.5 & 0.06 & 95.1 & 0.07 & 97.0 & 0.00 & 6.1 & 0.22 & 94.6 & 0.24 & 96.5 \\
    & M2 & -0.00 & 1.4 & 0.06 & 92.4 & 0.06 & 94.0 & 0.00 & 5.6 & 0.19 & 92.6 & 0.21 & 96.2 \\
  \midrule
  \multicolumn{14}{c}{$\Sigma$ Setting 2: $\Sigma_{\mathrm{ST}}$; SNR $\lambda = 1$}\\
    50 & M1 & -0.00 & 1.7 & 0.05 & 87.2 & 0.06 & 92.4 & 0.00 & 5.4 & 0.18 & 89.4 & 0.23 & 95.1 \\
    & M2 & -0.00 & 1.6 & 0.05 & 86.4 & 0.06 & 91.6 & 0.00 & 5.3 & 0.16 & 87.2 & 0.21 & 92.9 \\
    100 & M1 & -0.00 & 1.3 & 0.04 & 89.4 & 0.05 & 92.4 & -0.00 & 3.8 & 0.14 & 92.9 & 0.17 & 96.5 \\
    & M2 & -0.00 & 1.2 & 0.04 & 86.4 & 0.04 & 91.3 & -0.00 & 3.8 & 0.13 & 91.8 & 0.14 & 93.5 \\
    200 & M1 & 0.00 & 1.0 & 0.04 & 92.9 & 0.04 & 97.0 & 0.00 & 2.9 & 0.12 & 94.3 & 0.14 & 96.7 \\
    & M2 & 0.00 & 0.9 & 0.03 & 90.7 & 0.04 & 94.6 & 0.00 & 2.6 & 0.11 & 94.0 & 0.12 & 95.6 \\
    400 & M1 & -0.00 & 0.7 & 0.03 & 96.5 & 0.03 & 97.5 & -0.00 & 2.6 & 0.10 & 94.6 & 0.12 & 97.0 \\
    & M2 & -0.00 & 0.7 & 0.03 & 95.1 & 0.03 & 96.5 & -0.00 & 2.5 & 0.09 & 94.3 & 0.10 & 95.6 \\
  \bottomrule
\end{tabular}
\end{table}


\bibliographystyle{chicago}
\bibliography{fpref}

\label{lastpage_main}

\end{document}